\documentclass[12pt, draftclsnofoot, onecolumn]{IEEEtran}
\pdfoutput=1

\IEEEoverridecommandlockouts

\normalsize
\usepackage{cite}
\usepackage[normalem]{ulem}
\usepackage{color}
 \ifCLASSINFOpdf
   \usepackage[pdftex]{graphicx}
   \DeclareGraphicsExtensions{.pdf}
 \else
   \usepackage[dvips]{graphicx}
   \DeclareGraphicsExtensions{.eps}
 \fi

 \ifCLASSOPTIONcompsoc
	\usepackage[caption=false,font=footnotesize,labelfont=sf,textfont=sf]{subfig}
\else
	\usepackage[caption=false,font=footnotesize]{subfig}
\fi
\RequirePackage{lineno}

\usepackage{color}
\usepackage[usenames]{xcolor}

\usepackage{arydshln,leftidx,mathtools}
\usepackage{booktabs}
\usepackage[inline]{enumitem}
\usepackage{fixltx2e}

\usepackage{amsmath,amssymb,amsfonts,bbm}
\usepackage{arydshln,leftidx,mathtools}
\usepackage[backgroundcolor=cyan,bordercolor=gray,textsize=small,textwidth=5.8cm]{todonotes}

\usepackage{url}

\usepackage{algorithm}
\usepackage{algpseudocode}
\usepackage[flushleft]{threeparttable}
    
\ifCLASSINFOpdf
   \DeclareGraphicsExtensions{.pdf}
\else
   \DeclareGraphicsExtensions{.eps}
\fi

\DeclareMathOperator*{\rank}{rank}

\newtheorem{proposition}{Proposition}

\newcommand{\blackcircle}{\tikz[baseline]{\draw[black,solid,line width = 0.7pt](2.7mm, 1mm) circle (0.8mm)}}
\newcommand{\bluesquare}{\tikz[baseline]{\draw[blue,solid,line width = 0.7pt] (0.2mm,0.2mm) -- (1.8mm,0.2mm) -- (1.8mm,1.8mm) -- (0.2mm,1.8mm) -- (0.2mm,0.2mm)}}
\newcommand{\redcross}{\tikz[baseline]{\draw[red,solid,line width = 0.7pt](0.2mm,0.2mm) -- (1.8mm,1.8mm) -- (1mm,1mm) -- (1.8mm,0.2mm) -- (0.2mm,1.8mm)}}
\newcommand{\magentadiamond}{\tikz[baseline]{\draw[magenta,solid,line width = 0.7pt](0mm,1mm) -- (0.7mm,2mm) -- (1.4mm,1mm) -- (0.7mm,0mm) -- (0mm, 1mm)}}
\newcommand{\purpletriangle}{\tikz[baseline]{\draw[violet,solid,line width = 0.7pt](0.2mm,0.2mm) -- (0.2mm,1.8mm) -- (1.3mm,1mm) -- (0.2mm, 0.2mm) -- (0.2mm,1.8mm)}}
\newcommand{\grayplus}{\tikz[baseline]{\draw[gray,solid,line width = 0.7pt](0mm,1mm) -- (1.8mm,1mm) -- (0.9mm,1mm) -- (0.9mm,1.8mm) -- (0.9mm,0.2mm)}}

\newcommand{\itemcircle}{\tikz[baseline]{\draw[black,solid,fill=black,line width = 0.1mm](2.7mm, 1mm) circle (0.7mm)}}

\newlist{myitemize}{itemize}{3}
\setlist[myitemize]{noitemsep, topsep=0pt}
\setlist[myitemize,1]{label=\protect\itemcircle ,leftmargin=1.0in}
\setlist[myitemize,2]{label=$\rightarrow$,leftmargin=1em}
\setlist[myitemize,3]{label=$\diamond$}

\def\BibTeX{{\rm B\kern-.05em{\sc i\kern-.025em b}\kern-.08em
    T\kern-.1667em\lower.7ex\hbox{E}\kern-.125emX}}

\begin{document}

\title{
Coherent Multi-antenna Receiver for BPSK-modulated Ambient Backscatter Tags
}

\author{Xiyu Wang,
H{\"u}seyin~Yi\u{g}itler, Ruifeng Duan, Estifanos Yohannes Menta and Riku~J\"{a}ntti
\thanks{
The authors are with the Department of Communications and Networking, Aalto University, Espoo, 02150 Finland.
(e-mail: firstname.surname@aalto.fi)
}
}

\maketitle

\begin{abstract}

Ambient Backscatter Communication (AmBC) is an emerging communication technology that can enable green Internet-of-Things deployments. The widespread acceptance of this paradigm is limited by low Signal-to-Interference-Plus-Noise Ratio (SINR) of the signal impinging on the receiver antenna due to the strong direct path interference and unknown ambient signal. The adverse impact of these two factors can be mitigated by using non-coherent multi-antenna receivers, which is known to require higher SINR to reach Bit-Error-Rate (BER) performance of coherent
receivers. However, in literature, coherent receivers for AmBC systems are little-studied because of unknown ambient signal, unknown location of AmBC tags, and varying channel conditions. In this paper, a coherent multi-antenna receiver, which does not require a prior information of the ambient signal, for decoding Binary-Phase-shift-Keying (BPSK) modulated signal is presented. The performance of the proposed receiver is compared with the ideal coherent receiver that has a perfect phase information, and also with the performance of non-coherent receiver, which assumes distributions for ambient signal and phase offset caused by excess length of the backscatter path.  Comparative simulation results show the designed receiver can achieve the same BER-performance of the ideal coherent receiver with 1-dB more SINR, which corresponds to 5-dB or more gain with respect to non-coherent reception of On-Off-Keying modulated signals. Variation of the detection performance with the tag location shows that the coverage area is in the close vicinity of the transmitter and a larger region around the receiver, which is consistent with the theoretical results.

\end{abstract} 


\section{Introduction}
%
Recent advances in computing and communication technologies have enabled data exchanges among different devices, known as Internet-of-Things. The widespread deployments of IoT networks are inevitably limited by the scarcity of the communication spectrum and the power consumption of the devices. Although traditional communication technologies fall short for enabling massive IoT deployments, the recently emerging Ambient Backscatter Communication (AmBC) paradigm has the potential to provide a solution for both of the problems~\cite{Liu2013}. Since the technology requires neither power-hungry amplifiers nor a dedicated reader which generates specific carrier signal for sensors, AmBC realizes ultra low-power wireless communication. Significant bandwidth efficiency enhancement can also be obtained by using the spectrum allocated for a legacy system~\cite{Duan2017}. Possessing these features, AmBC has the potential to become an important component for realizing a sustainable IoT ecosystem.

In a typical AmBC system, a passive Backscatter Device (BD), often referred to as \emph{tag}, operates using the harvested ambient energy~\cite{ZhouTradeoff2013}. It transmits its information by modulating directly on top of the ambient Radio Frequency (RF) signal, such as cellular network~\cite{Parks2014}, WiFi~\cite{HWang2019}, and Digital Video Broadcasting - Terrestrial (DVB-T)~\cite{Vyas2013}. The state-of-the-art tags perform the On-Off-Keying (OOK) modulation by either \emph{backscattering} or \emph{absorbing} the ambient RF signal. The altered signal impinges on the receiver antenna together with the ambient RF signal. Then, the receiver recovers the transmitted tag information from the composite signal. This is usually done by the non-coherent detection as it does not require phase synchronization~\cite{Huynh2018}.
Detecting the backscatter information at the receiver, however, is limited by two properties of the AmBC system. First, the backscatter signal suffers from a strong Direct Path Interference (DPI) \cite{Duan2020Mag}. This results from the keyhole channel property of the backscatter path which causes a substantial signal strength loss.
And second, due to the lack of cooperation between the legacy system and the backscatter system, the receiver has little information about the ambient RF signal.  
These two properties of the backscatter signal particularly hampers its Signal-to-Interference-Plus-Noise  Ratio  (SINR) which limits the Bit-Error-Rate (BER) performance of AmBC systems.

Available solutions for addressing the strong DPI and unknown ambient signal are either eliminating legacy signal using complex signal processing techniques~\cite{Yang2018Cooperative}, or mitigating their impact using any of the direction~\cite{Duan2019} or spectral~\cite{ElMossallamy2019} differences between two signals. Among these solutions, direction difference is realized by multi-antenna receivers, which address these problems without any information on the ambient signal or on the channels, and they do not put any specific requirements on AmBC system setups. Since a multi-antenna receiver provides higher degree-of-freedom for implementation compared with other techniques, it is an ideal candidate for improving the observed SINR.

Another technique to improve detection performance of the receivers is to use Binary Phase Shift Keying (BPSK) modulation at BD. Compared with the commonly adopted OOK demodulators, BPSK demodulators can achieve the same BER-performance with 6 dB less SNR~\cite[Chapter~4]{Proakis2001}. In order to achieve all the SNR gain, BPSK demodulator should be implemented as a coherent receiver. Such a coherent receiver for AmBC systems requires a complex phase synchronization method since phase of the ambient signal and the channels, and the phase offset caused by the excess path length of the backscatter path compared with the direct path are not known at the receiver. Although coherent receiver should take into account all of these phases, multi-antenna receiver can be exploited to avoid the first two phases leaving only the last one to be estimated for realizing a coherent BPSK demodulator~\cite{Duan2019}.  
However, the low SINR of tag signal degrades the performance of standard phases estimation techniques, and thus more complex estimation methods should be adopted. Therefore, the BER-performance of AmBC systems, which can be translated to communication range or achievable rate, is improved by using BPSK modulation at BD and coherent demodulation at the multi-antenna receiver -- a solution has not been investigated in AmBC literature.

In this paper, we introduce a complete AmBC architecture for realizing the coherent reception of BPSK-modulated tag signal using multi-antenna receiver. We show that a three-state tag, which has an absorbing state and two states for BPSK, is needed for mitigating the DPI and for synchronizing the phase. Thereafter, we design a multi-antenna receiver architecture for retrieving AmBC signal. 
The receiver uses two beamformers to mitigate the DPI and unknown ambient signal, and utilizes a basic classification algorithm, Logistic Regression (LR), for demodulating the tag signal. The LR algorithm learns phase offset pattern from a training sequence and predicts the remaining transmitted bits based on the learned pattern. The primary contributions of this paper are as follows.
\begin{myitemize}[wide=0.1mm, leftmargin=1mm]
\item We formulate and solve the problem of coherent reception at a multi-antenna receiver of BPSK-modulated tag signal by taking the phase offset caused by excess backscatter path length into account. 

\item We design a coherent receiver architecture after deriving the sufficient statistic from the Maximum A Posterior (MAP) criterion, which does not require any prior information on the ambient signal or on the channels, and uses a simple classification algorithm to learn the pattern of phase offset. 

\item We derive the closed-form error probability of coherent receiver with ideal knowledge of phase offset to compare its BER-performance with the designed receiver performance. The results suggest that the designed receiver achieves the same BER-performance of the ideal coherent receiver when the signal has a maximum 1-dB more SINR, which corresponds to 5-dB or more gain with respect to non-coherent reception of OOK-modulated tag signal. 

\item We also derive the closed-form error probability of non-coherent receiver, which takes the form of energy detector, for given distributions of ambient signal and the phase offset. This receiver takes an energy-detector form, and it only works with OOK-modulated tag signal after canceling the DPI.

\item The developments of this work suggest that there are two key parameters affecting the BER-performance of AmBC receivers: the SNR of legacy system and the location of the tag. The latter parameter also dictates the coverage of an AmBC deployment, and is used for visualizing a spatial variation of Symbol Error Rate (SER). The results indicate that an acceptable performance is achieved when the SNR of legacy system is high and/or when the tag is in a close vicinity of the transmitter or in a large region ($\sim$ 20 wavelengths) around the receiver excluding the null beam of the receiver antenna array. 
\end{myitemize}

The rest of the paper is organized as follows. In Section~\ref{sec:related_work}, related studies are reviewed, and the notations used throughout the paper are introduced. The system model is outlined in Section \ref{sec:systemModel}.
In Section~\ref{sec:Method}, the designed receiver architecture is presented.  Theoretical error probabilities for the non-coherent receiver and the coherent receiver with known phase offset are derived in Section~\ref{sec:conventional_receivers}. 
In Section~\ref{sec:results}, the simulation results are presented, and important findings are discussed. Finally, conclusions are drawn in Section~\ref{sec:conclusion}.

\section{Background}\label{sec:related_work}

\subsection{Related work}
Our aim is to design a coherent receiver that achieves the SNR gain of BPSK-modulated tag signal in AmBC systems. We first provide a literature review about available solutions for mitigating the strong DPI and the unknown ambient signal. Then, we review existing phase synchronization methods for coherent receivers. Finally, we discuss other related techniques for improving the AmBC performance.

The DPI is the primary factor that causes low SINR backscatter signal regardless of the tag modulation. As discussed in~\cite{Duan2020Mag, Xiyu2019}, the backscatter path can be 30 dB weaker than that of the direct path when the tag is 3 meters away from the Rx.
Several DPI cancellation methods exist for different deployment scenarios. The DPI is mitigated using Successive Interference Cancellation (SIC) by jointly decoding ambient and backscatter signals when the channel coefficients are available at the receiver~\cite{Yang2018Cooperative}. For Orthogonal Frequency-Division Multiplexing (OFDM) ambient systems, tags may shift the frequency of the ambient signal to the guard bands between OFDM symbols~\cite{ElMossallamy2019} so that DPI can be eliminated by filtering, or they may operate on the cyclic prefix bands of OFDM systems~\cite{Yang2018OFDM}. 
Another way of mitigating DPI is to utilize multiple antennas at the receiver. The works \cite{Huayan2019} and \cite{Duan2019} have exploited spatial diversity in order to separate the backscatter path and the direct path from each other, which exempts the AmBC receiver from working with special ambient signal or from knowing channel state information. Different than~\cite{Huayan2019,Duan2019}, which are built upon non-coherent receivers, in this work, we present a multi-antenna receiver to coherently decode BPSK-modulated tag signal.

The unknown ambient signal can be eliminated by jointly estimating it along with the tag signal when the channels are known~\cite{Yang2018Cooperative}.  Such an estimation requires additional cooperation between the AmBC system and the legacy system, and thus have limited generality. AmBC receivers may avoid the necessity of tracking the unknown ambient signal by implementing a non-coherent receiver~\cite{Parks2014,Liu2013,Zhang2019,Yang2018Cooperative,Yang2018OFDM,ElMossallamy2019, Huayan2019, Qian2019, Darsena2019}.  By considering OOK and differential BPSK modulation, several works realize a non-coherent demodulator as an energy detector and maximum likelihood detector. BPSK modulation is also used with non-coherent receivers, however, it has been reported that severe error floor problem occurs if DPI is not canceled~\cite{Qian2019}. When DPI is canceled, BPSK-modulated tag signal cannot be decoded at the receiver. Different than these works, in this paper, we use a two-stage beamforming to mitigate the impact of unknown ambient signal on the receiver performance.

Designing a coherent receiver requires a complex phase synchronization because of an unknown phase offset caused by the propagation of the tag signal and unknown channel states. A coherent receiver and a partial coherent receiver are proposed in work~\cite{Vougioukas2019}, where channels are estimated by sending preambles from the transmitter and the tag. 
The phase offset is visualized in work~\cite{Zhao2019} and calibration techniques are proposed using preambles of the legacy system. These methods also need further cooperation between two systems, and thus have limited practical use. Although~\cite{Ma2018}  proposes a blind channel estimation using Expectation Maximization (EM) algorithm, the method needs prior information about the ambient signal constellation. In this paper, we propose a Machine Learning (ML)-assisted phase synchronization which neither require cooperation between the legacy and the AmBC systems nor any prior knowledge of the ambient signal.

Our work is related to several works on the demodulation with machine learning algorithms, which improves the BER-performance of AmBC systems by predicting the transmitted signal after learning the received signal patterns from training sequences~\cite{HWang2019, Zhang2019}. 
The work~\cite{Zhang2019} applies  an EM assisted method to retrieve the OOK-modulated tag signal, and \cite{HWang2019} extracts a unique slope feature of the received WiFi signal. Nevertheless, these methods require the receiver to know the constellation of the legacy system. Our method does not depend on any information of the legacy system, and thus is more general than the aforementioned works.

In this work, we also use coding to improve the successful tag signal recovery rate. Similar coding techniques have already been used in works~\cite{Parks2014} and special waveform designs have been investigated in works~\cite{Yang2018OFDM, ElMossallamy2019, Nguyen2019OFDM} to enhance the communication performance of battery-free tags. In sequel, we study the impact of well-known coding methods on the receiver performance.

Finally, this paper is an extension of our previous work~\cite{Xiyu2019} which presents a ML-assisted receiver and evaluates its BER-performance. In this paper, we rigorously derive the proposed receiver from the MAP criterion, and analyze a coherent receiver with perfect phase information and a non-coherent receiver. Further, we present a three-state AmBC tag modulator. We carry out frequency independent simulations to compare the performance of the designed receiver with two receivers. We illustrate variation of the receiver performance with the tag location to suggest a coverage area for the tag. 

\subsection{Notations}
Throughout the paper, scalars are denoted by normal font letters $a$, vectors and matrices are represented by lower-case $\boldsymbol{a}$ and upper-case $\boldsymbol{A}$ boldface letters, respectively. Complex valued scalars are assumed, and $\mathrm{Re}\{a\}$ and $\mathrm{Im}\{a\}$ denote the real and the imaginary parts of a scalar $a$, respectively.
The Euclidean norm of a vector $\boldsymbol{a}$ is denoted by $\|\boldsymbol{a}\|$. The $n\times n$ identity matrix is $\boldsymbol{I}_n$, and the subscript $n$ may be omitted sometimes for notational convenience. The conjugate-transpose, conjugate and transpose of a matrix $\boldsymbol{A}$ are $\boldsymbol{A}^H$, $\boldsymbol{A}^*$ and $\boldsymbol{A}^T$ respectively, and $\rank (\boldsymbol{A})$ is its rank. We use $\mathcal{CN}(\boldsymbol{m}, \boldsymbol{\Sigma})$ to denote the circularly symmetric complex Gaussian variable with mean $\boldsymbol{m}$ and covariance matrix $\boldsymbol{\Sigma}$.  The statistical expectation is $\mathrm{E}\{\cdot\}$, variance is $\mathrm{Var}\{\cdot\}$, and probability of an event is $\mathbb{P}\{\cdot\}$.  The imaginary number is  $j = \sqrt{-1}$.

\section{Problem Formulation}\label{sec:systemModel}
\begin{figure}[!t]
    \centering
    \includegraphics[width=0.47\textwidth]{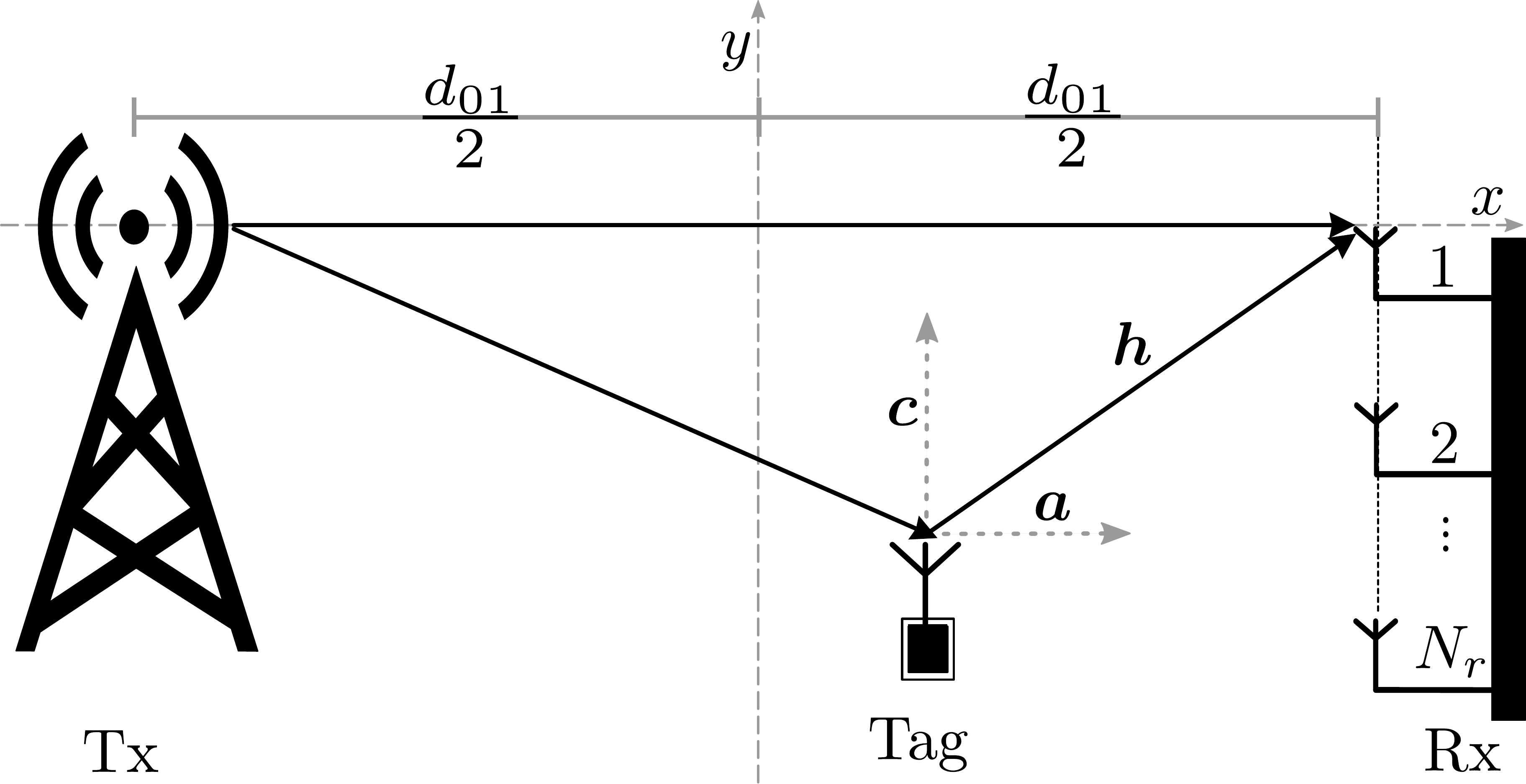}
    \caption{An illustration of AmBC system with a multi-antenna receiver and a single-antenna tag in a two dimensional Euclidean space. The line connecting the Tx antenna and the reference Rx antenna is $x$-axis. The middle point of two points is the origin. The ambient RF signal transmitted from the Tx propagates to the Rx through direction $\boldsymbol{a}$. The signal backscattered by the tag propagates to the Rx through direction $\boldsymbol{h}$. The direction $\boldsymbol{h}$ can be decomposed by an orthonormal basis $\boldsymbol{a}$ and $\boldsymbol{c}$. }
     \label{fig:systemModel}
\end{figure}

In this paper, we consider a basic AmBC system,  which consists of a backscatter device (Tag), a separated legacy ambient source (Tx) and an AmBC receiver (Rx) with $N_r$ antennas as shown in Fig.~\ref{fig:systemModel}. In the illustrated scenario, the passive tag modulates its own information onto the ambient signal, and the Rx receives both ambient signal and backscatter signal. In this section, we present a system model for this scenario, define the general terms used throughout the paper, and give practical assumptions on the wireless channels, the ambient signal and the tag signal.

\subsection{Channel model}
In the remaining of the paper, without loss of generality, we consider a two dimensional Euclidean space with Cartesian reference frame shown in Fig.~\ref{fig:systemModel}. The first receiver antenna out of $N_r$ elements is selected as the reference antenna. The line connecting the Tx and the reference antenna is set to be $x$-axis and middle point of this line segment is set as the origin of the reference frame.  The position of the $l$th Rx antenna, $l = 1, \cdots, N_r$, the position of the Tx and the position of the tag are denoted by $\boldsymbol{p}_{rl}$, $\boldsymbol{p}_t$ and $\boldsymbol{p}$, respectively. Then, the distance between the Tx and the $l$th Rx antenna is $d_{0l} = \|\boldsymbol{p}_t - \boldsymbol{p}_{rl}\|$, the distance between the tag and the $l$th Rx antenna is $d_{1l} = \|\boldsymbol{p} - \boldsymbol{p}_{rl}\|$ and the distance between the Tx and the tag is $d_2 = \|\boldsymbol{p} - \boldsymbol{p}_t\|$.

Let the vectors $\hat{\boldsymbol{a}} = [\hat{a}_1, \cdots, \hat{a}_{N_r}]$ and $\hat{\boldsymbol{h}} = [\hat{h}_1, \cdots, \hat{h}_{N_r}]$ represent channel gains of the direct path and the backscatter path seen at the Rx, respectively. In the simplest form, the channel gains of the direct path of the $l$th Rx antenna $\hat{a}_l$ and of the backscatter path of the $l$th Rx antenna $\hat{h}_l$ are 
\begin{equation}
\begin{aligned}
     \hat{a}_l  &=\left(\frac{ \lambda}{4 \pi d_{0l}}\right)^2\exp\left( \frac{j 2 \pi d_{0l}}{\lambda} \right), \\
     \hat{h}_l &= \left(\frac{ \lambda}{4 \pi d_{1l}}\right)^2\left(\frac{ \lambda}{4 \pi d_2}\right)^2 \exp\left( \frac{j 2 \pi (d_{1l}+d_2)}{\lambda} \right),
     \end{aligned}
     \label{eq:channel_gains}
\end{equation}
where $f_c$ is the carrier frequency, $\lambda =f_c/c_0$ is the carrier wavelength and $c_0$ is the free-space electromagnetic propagation speed. The channel gains of two paths are normalized with the channel gain of direct path so that
\begin{equation*}
    \tilde{\boldsymbol{a}} = \frac{\hat{\boldsymbol{a}}}{\|\hat{\boldsymbol{a}}\|}, \quad \tilde{\boldsymbol{h}} = \frac{\hat{\boldsymbol{h}}}{\|\hat{\boldsymbol{a}}\|}  .
\end{equation*}
Although the channels are different when the multi-path fading is considered, its impact is not on the directions, and thus, not affecting the following derivations.

\subsection{Tag signal}
As can be seen from Eq.~\eqref{eq:channel_gains}, the term $(\lambda/(4\pi))^2$ results in an extensive power loss for the backscatter path such that its effective SNR is still relatively low even after the DPI cancellation. We use two techniques at the tag in order to improve the effective SNR of the backscatter signal. First, the tag adopts BPSK modulation by altering the antenna impedance to switch the phase of incident RF signal. 
Second, we utilize orthogonal Hadamard code and non-orthogonal Simplex code, which are commonly used due to their easy implementation~\cite{Gallager2008}. Hadamard code is able to correct many errors by sacrificing the efficiency, which makes it a good candidate for the noise-corrupted backscatter channels. The Simplex code achieves the same performance as Hadamard code with one dimension less codewords~\cite{Gallager2008}.

At the tag, a data-bit sequence $\boldsymbol{B}$ of $\{0,1\}$ is segmented into $P$ length-$(r+1)$ tuples. The length-$(r+1)$ tuple is encoded to a length-$n=2^{r+1}$ codeword by using the generator matrix, where $r$ is called the code order. The Hadamard code generator matrix $G_{H,r}$ of order $r$ can be expressed as~\cite{Yue2007}:
\begin{equation*}
    G_{H, r} = \begin{bmatrix}G_{H, r-1} &  J_{r-1}\oplus G_{H, r-1} \\ \begin{matrix} 0 & 0 \cdots & 0\end{matrix}, & \begin{matrix} 1 & 1 \cdots & 1\end{matrix}\end{bmatrix}, \quad r\geq 2,
\end{equation*}
where $G_{H, r}$ is a $(r+1)\times2^{r+1}$ matrix, $J_{r-1}$ and $G_{H,r-1}$ have the same size, $\oplus$ is the binary addition operator, and 
\begin{equation*}
    G_{H,1} = \begin{bmatrix}1 & 0 & 0 & 1 \\ 0 & 1 & 0 & 1 \end{bmatrix}, \quad J_{r} = \begin{bsmallmatrix}1 & 1 & \cdots & 1 \\ 0 & 0 & \cdots & 0 \\ \vdots & \vdots & \ddots & \vdots \\0 & 0 & \cdots & 0 \end{bsmallmatrix}.
\end{equation*}
Correspondingly, the generator matrix for the Simplex code $G_{S,r}$, a $(r+1) \times (2^{r+1}-1)$ matrix,  can be obtained by removing the all-zero column from $G_{H,r}$. 
Denoting the $\ell$th tuple, $\ell = 1, \cdots, P$, as  $\boldsymbol{b}_{r,
\ell} = \begin{bmatrix}b_{0,\ell} & b_{1,\ell} & \cdots & b_{r,\ell}\end{bmatrix}$. The resulting codeword  $\tilde{x}_{\ell}$, also referred to as \emph{symbol}, can be obtained from $\boldsymbol{b}_{r,\ell}$ as 
\begin{equation*}
    \boldsymbol{\tilde{x}}_{\ell} = \mathcal{G}(\boldsymbol{b}_{r,\ell} \oplus G_{r}),
    \label{eq:encoder}
\end{equation*}
where $G_r \in \{G_{H,r}, G_{S,r}\}$ and $\mathcal{G}(\cdot)$ represents the mapping  $0 \rightarrow -1, 1 \rightarrow 1$.

\begin{figure}[!tb]
	\centering
	\setlength{\tabcolsep}{0pt}
	\begin{tabular}{c}
		\subfloat[]{\includegraphics[width=0.48\textwidth]{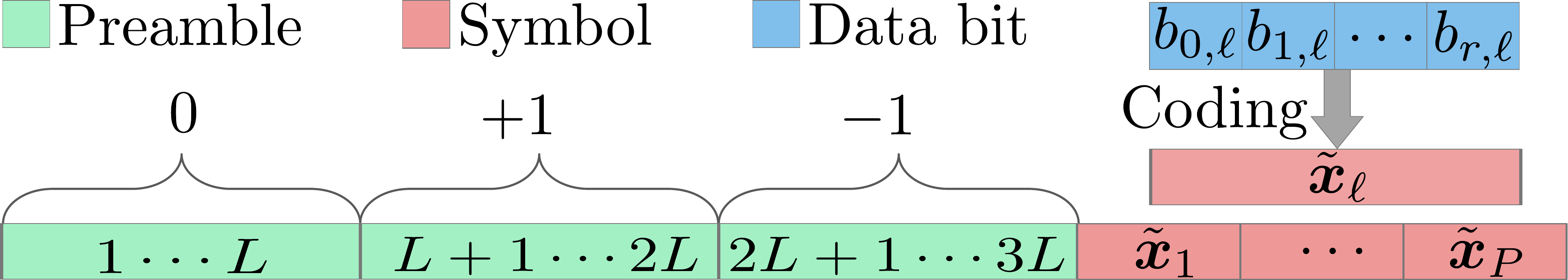}
			 \label{fig:duration}} \\
		\subfloat[]{\includegraphics[width=0.48\textwidth]{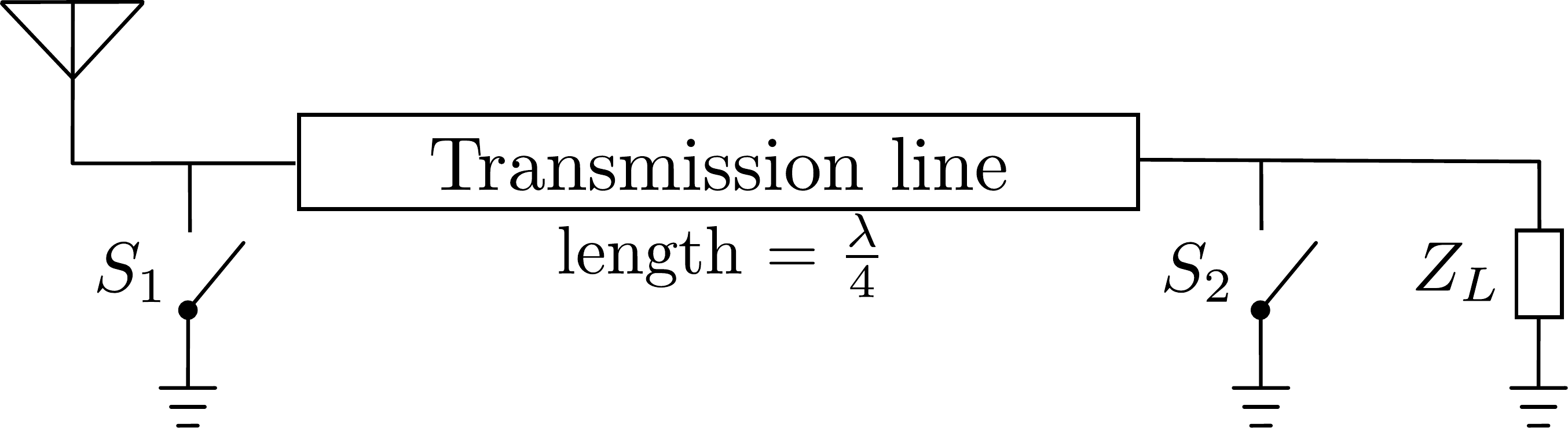} 
			\label{fig:tag_description}} 
	\end{tabular}
\caption{In (a), one tag frame over the channel coherence time consisting of  3 length-$L$ preambles and $P$ length-$n$ symbols. Preambles are associated with '0', '+1' and '-1', respectively. One symbol is encoded from a length-$r+1$ data-bit tuple.   In (b), illustration of the three-state modulator at the tag. When $S_1$ is closed and $S_2$ is open, the ambient signal is reflect back. When $S_2$ is closed and $S_1$ is open, the ambient signal is reflected back with an extra $\pi/2$ radians phase difference. When $S_1$ and $S_2$ are both open, it matches absorbed state, i.e., $x=0$. }
\end{figure}

The transmitted frame within one channel coherence time\footnote{We assume that the channel gains stay the same within one frame which implies that the channel coherence time is longer than one frame duration. This assumption can be easily fulfilled in slowly varying environments.} is shown in Fig. \ref{fig:duration}. The transmitted $P$ symbols, which are composed of $\{-1, 1\}$, are denoted as $\boldsymbol{X} = [\boldsymbol{\tilde{x}}_1,  \cdots, \boldsymbol{\tilde{x}}_P]$. Before transmitting $\boldsymbol{X}$, three length-$L$ preambles of only 0, only +1 or only -1 are prepended.  
Hence, a frame of the AmBC tag signal of length $N = 3L+nP$ is represented as $ \begin{bmatrix} x[1] & \cdots &  x[3L]  &\boldsymbol{X} \end{bmatrix}$ where $x[1]=\cdots=x[L]=0, x[L+1]=\cdots=x[2L]=+1, x[2L+1]=\cdots=x[3L]=-1$.  

The frame design requires a tag modulator that can be switched to any of three states. An implemetation of such tag is shown in Fig. \ref{fig:tag_description}, which consists of two single-pole switches $S_1$ and $S_2$ and one $Z_L$ impedence load that is equal to the conjugate of the antenna impedance. In principle, backscattering is achieved by altering the load impedance to change the reflection coefficient~\cite{Talla2013}. Specifically, when two switches are open, energy is absorbed by $Z_L$, and $x=0$. When either of switches is closed, the load impedance is zero such that all the signal are reflected. Since there is an extra transmission line with length $\lambda/4$ , the reflected signals with respect to two switches have $\pi/2$ radians phase difference, which realizes $+1$ and $-1$ of the BPSK modulation. 

\subsection{Received signal at the Rx}
Let us denote the $i$th ambient signal transmitted from Tx within one channel coherence time as $\tilde{s}[i]\sim \mathcal{CN}(0, 1), \; \forall i= 1, \cdots, N$, and denote $\gamma$  the received SNR  of the ambient signal at the reference antenna. 
When the signal reaches to the Rx reference antenna, it experiences the direct path channel $\tilde{\boldsymbol{a}}$ so that it becomes
\begin{equation*}
    \sqrt{\gamma}\tilde{\boldsymbol{a}}\tilde{s}[i] = \sqrt{\gamma}\boldsymbol{a}e^{j\phi_0}\tilde{s}[i] ,
\end{equation*}
where $\phi_0$ is the phase shift caused by the propagation.
The ambient signal traversing along the backscatter path has an extra phase offset $\phi$ caused by the excess path length, which is given by 
\begin{equation*}
    \sqrt{\gamma}\tilde{\boldsymbol{h}}\tilde{s}[i]x[i] = \sqrt{\gamma}e^{j(\phi +\phi_0)}\boldsymbol{h}\tilde{s}[i]x[i]  ,
\end{equation*}
where $\phi$ is 
\begin{equation*}
     \phi = 2\pi \frac{d_{11} + d_2 - d_{01}}{\lambda}  .
\end{equation*}
Since we are not interested in $\tilde{s}[i]$, we use $s[i] = \tilde{s}[i]e^{j\phi_0}$ to denote the ambient signal for notational convenience.

\begin{figure}
    \centering
    \includegraphics[width=0.465
    \textwidth]{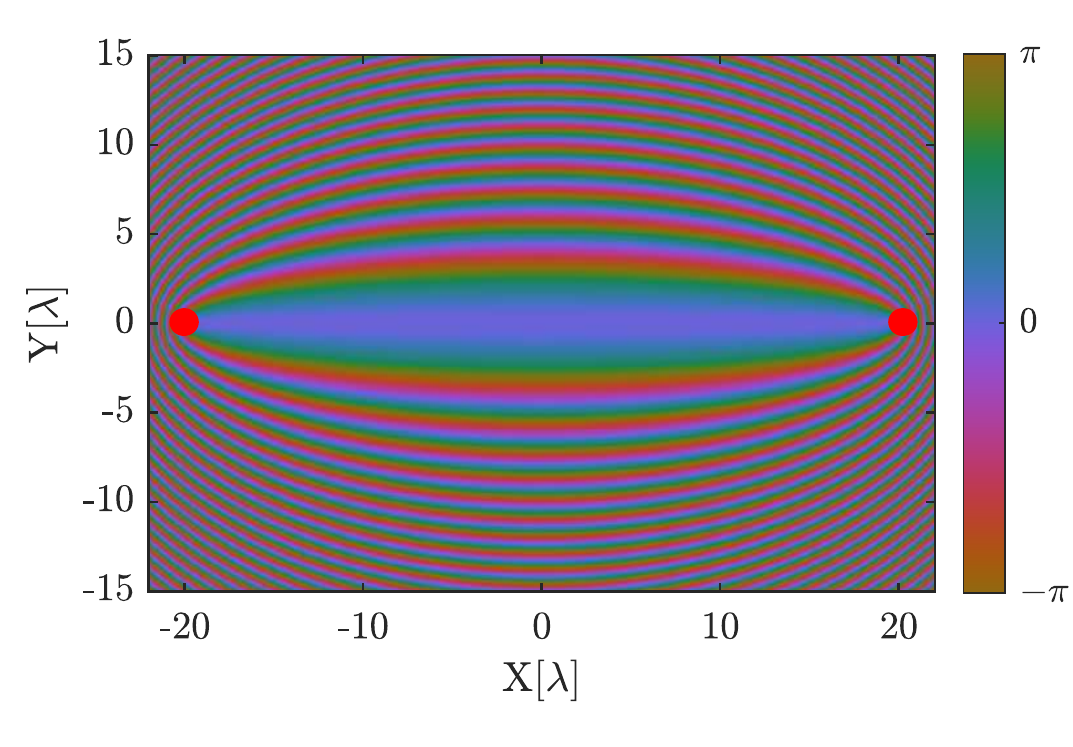}
    \caption{Variation of the phase offset $\phi$ as a function of tag location when direct path length $d_{01} = 40 \lambda$. The Tx antenna and the reference antenna of the Rx are marked as red circles.}
     \label{fig:phase_location}
\end{figure}

The variation of phase offset as a function of tag location is illustrated in Fig. \ref{fig:phase_location}, where the Tx antenna and the Rx reference antenna are marked by red circles. It can be observed that a small change in the tag location cause a large variation in $\phi$. The phase variation is the highest when the tag is close to either Tx or Rx. Therefore, ignoring phase offset while decoding the tag information, degrades the receiver performance.

The $i$th sample of additive white Gaussian noise in the received signal at $l$th Rx antenna  $\omega_l[i]$ is circularly symmetric Gaussian with unity variance so that  $\boldsymbol{\omega}[i] \in \mathbb{C}^{N_r}$ is $\boldsymbol{\omega}[i]\sim \mathcal{CN}(\boldsymbol{0}, \boldsymbol{I}_{N_r})$ which is independent of the ambient signal and the backscatter signal.  Then, the $i$th sample of received signal, $i = 1, \cdots, N$, is given by 
\begin{equation*}
\begin{aligned}
    \boldsymbol{y} [i] = \sqrt{\gamma}\big( \boldsymbol{a}  s[i] + e^{j \phi} \boldsymbol{h} s[i] x[i] \big)+ \boldsymbol{\omega}[i] .
\end{aligned}
\end{equation*} 
We call $\boldsymbol{a}$ and $\boldsymbol{h}$ the \emph{directions} of the direct path and the backscatter path. The power difference between two paths is defined as $\Delta = \|\boldsymbol{h}\|^2/\|\boldsymbol{a}\|^2 = \|\boldsymbol{h}\|^2$. 
And the direction $\boldsymbol{h}$ can be decomposed as
$\boldsymbol{h} = \eta_1 \boldsymbol{a} + \eta_2\boldsymbol{c}$ (see Fig.~\ref{fig:systemModel}), where 
\begin{equation*}
\boldsymbol{c} = \frac{\left(\boldsymbol{I} -\boldsymbol{a}\boldsymbol{a}^H \right) \boldsymbol{h}}{\|\left(\boldsymbol{I} -\boldsymbol{a}\boldsymbol{a}^H \right) \boldsymbol{h}\| },
\end{equation*}
such that $\boldsymbol{c}^H \boldsymbol{a} = 0$, and $\eta_1, \eta_2$ are projections of $\boldsymbol{h}$ onto $\boldsymbol{a}$ and $\boldsymbol{c}$ satisfying $0 \leq \eta_1, \eta_2 \leq 1$ and $|\eta_1|^2 + |\eta_2|^2 = \Delta$\footnote{The component of $\boldsymbol{h}$ on $\boldsymbol{c}$ is a real number by definition of $\boldsymbol{c}$.}. Hence, the received signal can be rewritten as 
\begin{equation}
\label{eq:transceiver}
\begin{aligned}
    \boldsymbol{y} [i] = \sqrt{\gamma} \big( \boldsymbol{a}  s[i] + (\eta_1 \boldsymbol{a} + \eta_2\boldsymbol{c})e^{j \phi} s[i] x[i] \big)+ \boldsymbol{\omega}[i] .
\end{aligned}
\end{equation}

Let us also define the received signal sample matrices corresponding to the preambles as
\begin{subequations}
\begin{gather*}
\boldsymbol{Y}_0 = [\boldsymbol{y}[1], \cdots, \boldsymbol{y}[L]] ,\\ 
\boldsymbol{Y}_{t+} = \left[\boldsymbol{y}[L+1], \cdots, \boldsymbol{y}[2L]\right], \\
\boldsymbol{Y}_{t-} = [\boldsymbol{y}[2L+1], \cdots, \boldsymbol{y}[3L]] , \\
\boldsymbol{Y}_{t} = [\boldsymbol{Y}_{t+}, \boldsymbol{Y}_{t-}].
\end{gather*}
\end{subequations}


\begin{figure*}
    \centering{
    \includegraphics[width=.95\textwidth, keepaspectratio]{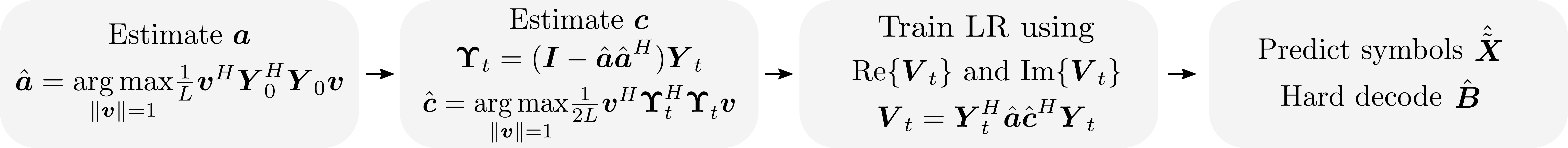}
    \caption{Flowchart of the proposed ML-assisted receiver: direction $\boldsymbol{a}$ is estimated from $\boldsymbol{Y}_0$ and direction $\boldsymbol{c}$ is estimated from the residual signal of $\boldsymbol{Y}_t$. Logistic regression (LR) algorithm is trained by taking $\mathrm{Re}\{\boldsymbol{V}_t\}$ and $\mathrm{Im}\{\boldsymbol{V}_t\}$ as two features. The data-bit sequence  $\hat{\boldsymbol{B}}$ is decoded from the predicted symbols $\hat{\tilde{\boldsymbol{X}}}$.} 
    \label{fig:flowchart}
    }
\end{figure*}

\section{Designed Receiver Architecture}\label{sec:Method}

In this section, the proposed ML-assisted receiver, summarized in Fig.~\ref{fig:flowchart}, is presented step-by-step.

\subsection{Direct path interference elimination}
The received signal $\boldsymbol{y}[i]$ is composed of both ambient signal and tag signal, which can be separated if the directions of these two paths are known, or estimated using the acquired samples. For this purpose, the received signal sample matrix corresponding to the first part of the preamble $\boldsymbol{Y}_0$ is used to estimate the direct path direction $\boldsymbol{a}$ 
%
by computing the eigenvector corresponding to the largest eigenvalue of sample covariance matrix of $\boldsymbol{Y}_0$, i.e.,
\begin{equation*}
    \hat{\boldsymbol{a}} = \underset{\|\boldsymbol{v}\| = 1}{\arg\max} \frac{1}{L}\boldsymbol{v}^H \boldsymbol{Y}_0^H\boldsymbol{Y}_0^{}\boldsymbol{v}. 
\end{equation*}
For the rest of the frame transmission, we eliminate the DPI by projecting the received signal into the orthogonal space of $\hat{\boldsymbol{a}}$, which yields a residual signal given by
\begin{equation}
\begin{aligned}
    \boldsymbol{r}[i] &= \left( \boldsymbol{I} -  \hat{\boldsymbol{a}}\hat{\boldsymbol{a}}^H \right) \boldsymbol{y}[i] \\
    &= \sqrt{\gamma} e^{j \phi} \eta_2 \boldsymbol{c} s[i]x[i] + \left( \boldsymbol{I} -  \hat{\boldsymbol{a}}\hat{\boldsymbol{a}}^H \right) \boldsymbol{\omega}[i]  ,
    \label{eq:resiSignal}
    \end{aligned}
\end{equation}
for $i = L+1, ~\cdots, N$.

\subsection{MAP receiver}
The residual signal given in Eq.~\eqref{eq:resiSignal} only contains the tag signal in the direction $\boldsymbol{c}$ (cf. Fig.~\ref{fig:systemModel}). This direction can be estimated using the same approach as in the previous subsection. For this purpose, let us denote $\Upsilon_t = (\boldsymbol{I} - \hat{\boldsymbol{a}}\hat{\boldsymbol{a}}^H)\boldsymbol{Y}_t$ as the residual signal matrix over the remaining two preambles. Then, the estimate of direction $\boldsymbol{c}$ is given by
\begin{equation*}
 \hat{\boldsymbol{c}} = \underset{\|\boldsymbol{v}\|=1}{\arg\max}\frac{1}{2L}\boldsymbol{v}^H\boldsymbol{\Upsilon}_t^H \boldsymbol{\Upsilon}_t^{}\boldsymbol{v}.  
\end{equation*}
Thereafter, the Rx performs the beamforming to combine the residual signal at each antenna, which yields the effective tag signal 
\begin{equation}
\begin{aligned}
    u[i] &= \hat{\boldsymbol{c}}^H \boldsymbol{r}[i] = \hat{\boldsymbol{c}}^H \boldsymbol{y}[i] \\
   & = \sqrt{\gamma} e^{j\phi} \eta_2 s[i]x[i] + \hat{\boldsymbol{c}}^H \boldsymbol{\omega}[i] ,
\end{aligned}
\label{eq:relevantSignal}
\end{equation}
where $\boldsymbol{c}^H \boldsymbol{\omega}[i] \sim \mathcal{CN}(0,1)$ is the projected noise, and $\gamma|\eta_2|^2$ is defined as the effective SNR. The following Proposition gives the testing statistics for an optimum receiver.

\begin{proposition}\label{prop:testing_statistic}
For the effective tag signal in Eq.~\eqref{eq:relevantSignal}, the testing statistic of the optimum receiver for tag signal reads as~\cite[Chapter~4]{Wozencraft1965}
\begin{equation}
\begin{aligned}
    \mathrm{Re}\{ e^{-j\phi} &\hat{s}^*[i] u[i]\} = \mathrm{Re}\{e^{-j\phi} \boldsymbol{y}^H[i] \hat{\boldsymbol{a}}\hat{\boldsymbol{c}}^H  \boldsymbol{y}[i]\}\\
    &=  \cos\phi \cdot \mathrm{Re}\{v[i]\} + \sin\phi \cdot \mathrm{Im}\{v[i]\} ,
\end{aligned}
\label{eq:dataset}
\end{equation}
where $\hat{s}[i] = \hat{\boldsymbol{a}}^H \boldsymbol{y}[i]$ and $v[i] = \hat{s}^*[i] u[i]$.
\end{proposition}
\begin{IEEEproof}
See Appendix \ref{appendix:testing_statistic}.
\end{IEEEproof}

The above result implies the sufficient statistic of the optimum receiver correlates the effective signal $u[i]$ with the partial estimate of ambient signal $\hat{s}[i]$.  It is of importance to notice that the testing statistic is affected by the phase offset $\phi$ such that both $\mathrm{Re}\{v[i]\}$ and $\mathrm{Im}\{v[i]\}$ affect its value. Hence, ignoring $\phi$ and only looking at the $\mathrm{Re}\{v[i]\}$ downgrades the BER-performance as shown in Section~\ref{sec:results}. An extreme case is when $\phi = \pi/2$ radians ($\mathrm{Re}\{v[i]\} = 0$), which yields the testing statistic to be $\mathrm{Im}\{ v[i]\}$, although the tag signal may still be decoded. However, in practice, it is challenging to compensate for the phase offset $\phi$ since the performance of well-known phase estimation/compensation methods for low SINR effective signal $u[i]$ are not acceptable, and thus $\phi$ must be compensated for by other means. 

\begin{figure}[t!]
\centering
\begin{tabular}{cc}
		\subfloat[]{\includegraphics[width=0.42\textwidth]{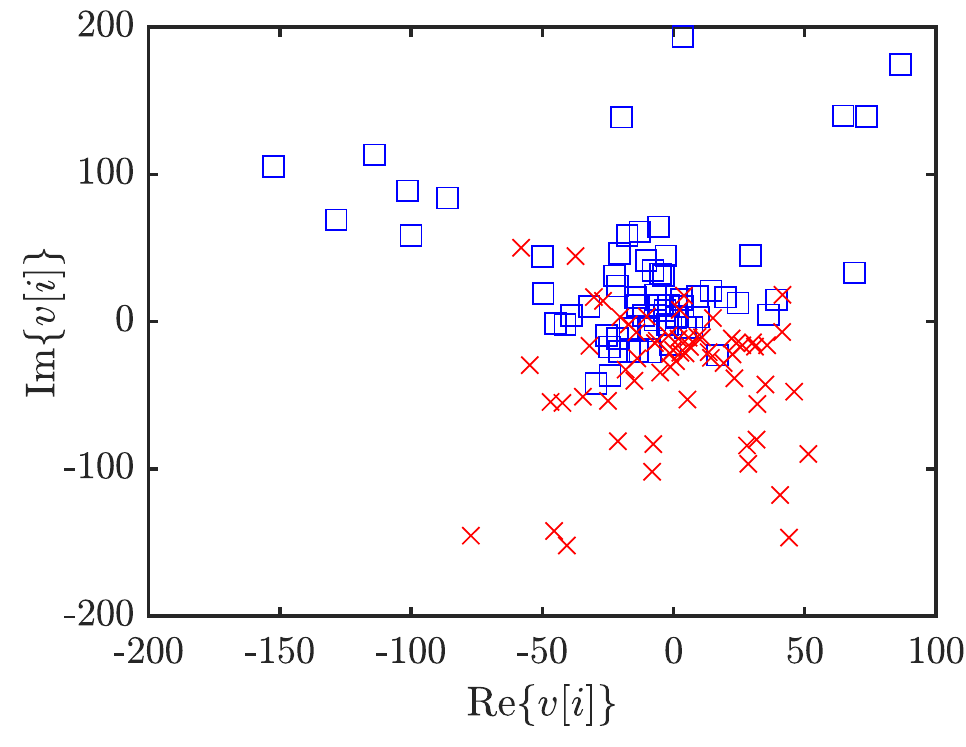}
			 \label{fig:training_data}} 
		\subfloat[]{\includegraphics[width=0.42\textwidth]{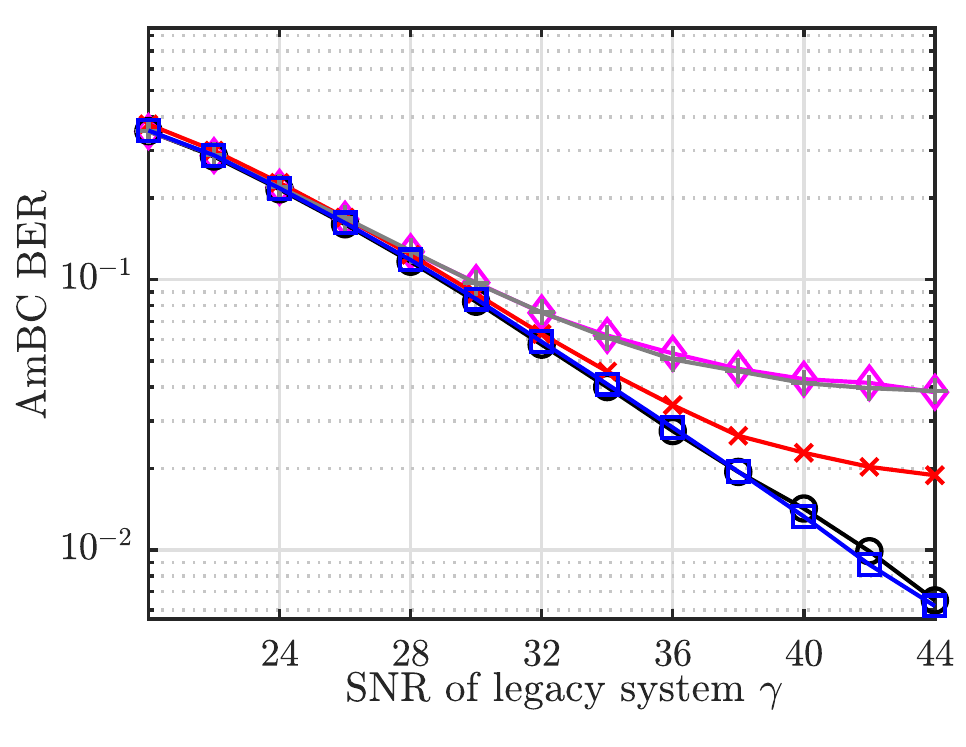} 
			\label{fig:algos}} 
\end{tabular}
\caption{In (a), visualization of training set when preamble size $L=64$, power difference between two paths $\Delta \approx -35$ dB and SNR of legacy system $\gamma = 28$ dB. The markers represent two classes: +1 (\protect\redcross) and -1 (\protect\bluesquare). In (b), variation of AmBC Bit-Error-Rate (BER) as a function of the SNR of legacy system $\gamma$ for $L = 64$ and $\Delta \approx -35$ dB with different markers for different machine learning classification algorithms: linear discriminant analysis (\protect\magentadiamond), least-squares-based classifier (\protect\grayplus), k-nearest-neighbors with $k=23$ (\protect\redcross), soft margin support vector machine (\protect\bluesquare) and logistic regression (\protect\blackcircle). }
\end{figure}

In Fig. \ref{fig:training_data}, the values of $[\mathrm{Re}\{v[i]\}, \mathrm{Im}\{v[i]\}]$ when $x=-1$ and $x=+1$ are shown\footnote{Note that positions of the clusters vary with channel conditions and the SNR of legacy system. The effective signal values shown in Fig.~\ref{fig:training_data} are acquired within one channel coherence time. }. As can be seen, the values fall into two clusters, implying that an ML classification algorithm can be used for learning the pattern of unknown phase offset. In the next subsection, we elaborate on ML classification algorithms which use $\mathrm{Re}\{v[i]\}$ and $\mathrm{Im}\{v[i]\}$ as their features.


\subsection{ML-based demodulation}
Variation of dataset $[\mathrm{Re}\{v[i]\}, \mathrm{Im}\{v[i]\}]$ instantiated in Fig.~\ref{fig:training_data} shows that there exist outliers and the dataset corresponding to $x=+1$ and $x=-1$ classes are overlapping because of the low SINR of the tag signal. As SINR increases, the overlapping area diminishes. However, it will not disappear according to the data distribution which we will analyze in the next section. Therefore, both linear and non-linear classifiers can be utilized for the purpose.

Common linear classifiers including Logistics Regression (LR), soft margin Support Vector Machine (SVM), Linear Discriminant Analysis (LDA), least-square-based classifier, and a simple non-linear classifier, k-nearest neighbor (kNN), are candidates for learning the pattern from the dataset. 
Among these algorithms, soft margin SVM can be configured by using a hyperparameter $C$ to control the weight of soft margin latent variables, which in turn defined how well it fits to the training data. Although SVM and LR have similar cost functions, LR casts the fitting problem using the Sigmoid function and looks at the probabilities of an observation being in either of the classes. The SVM has similar performance to LR when $C$ is adjusted. The least-squares-based classifier obtains the parameters by minimizing the prediction error, and LDA gets the parameters by maximizing the class separation. However, for predicting BPSK-modulated tag signal with equal probability, they are equivalent to each other \cite[Chapter~4]{Bishop}. The non-linear kNN classifies data intuitively by assigning it to the class which has the majority votes among $k$ selected nearest neighbors. The selection criterion for a classification method includes its performance, but also its computational requirements and the length of the training data must be taken into account. 

Let us denote $\boldsymbol{V}_t = \boldsymbol{Y}_t^H \hat{\boldsymbol{a}}\hat{\boldsymbol{c}}^H\boldsymbol{Y}_t^{}$ as testing statistics of two length-$L$ preambles, which are used as the training set to calculate the decision boundary of two classes. The rest of the transmitted bits are classified by comparing their testing statistics with the decision boundary. An example of the BER-performances of different ML algorithms is displayed in Fig.~\ref{fig:algos}. It shows that in low SNR region, these ML algorithms have almost the same performances. In high SNR region, soft margin SVM and LR still have similar performance and they outperform the others. As discussed above, LDA and least square classification also perform similarly, but compared with the SVM and LR, they lack robustness to outliers. Also, the assumption of LDA that observations of each class follow Gaussian distribution is not applicable to the studied case. On the other hand, kNN has acceptable performance as has been evaluated in our previous work \cite{Xiyu2019}. It can be highly accurate with a large training size and for an appropriate number of neighbors $k$. However, this requirement cannot be always satisfied since short preambles should be designed in order to save tag energy. Furthermore, large preamble size and $k$ also introduce high computational complexity and higher memory requirements. Consequently, logistic regression classifier is the most suitable for mitigating the impact of the unknown phase offset $\phi$.

\subsection{Decoder}
The output of the classifier, $P$ distorted symbols denoted by  $\hat{\tilde{\boldsymbol{X}}}= [\hat{\tilde{\boldsymbol{x}}}_1, \cdots, \hat{\tilde{\boldsymbol{x}}}_P]$, are then input into a hard-decision decoder to recover the data bits~\cite[Section~7.5]{Proakis2001}. In this paper, a conceptually simple minimum-distance decoding is adopted since the main focus is on demodulation. The decoding process for one received symbol is done by comparing it with $n$ possible transmitted codewords and selecting the one that has the lowest Hamming distance. This can be achieved by correlating $\hat{\boldsymbol{x}}_{\ell}, \ell = 1, \cdots, P$ with $n$ codewords and output the one with the largest correlation.  Finally, after retrieving all the data bits, the BER as well as Symbol Error Rate (SER) of AmBC system can be calculated.

\section{Conventional receivers}\label{sec:conventional_receivers}

In this section, two conventional receivers are elaborated on. The first one is the coherent receiver which requires knowledge of the phase offset or can compensate for it,  and the other is the non-coherent receiver which averages out the ambient signal and the phase offset. First, error probabilities of two receivers are given. Then, relative parameters associated with the detection performance are discussed.

\subsection{Coherent receiver for known phase offset}
The coherent receiver requires the knowledge of phase offset at the Rx. To analyze its performance, let us rewrite its sufficient statistic in Eq.~\eqref{eq:dataset}  as\footnote{In the remaining part of this paper, we will drop the time dependence of $\boldsymbol{y}$ since the detection of tag signal is based on single sample of $\boldsymbol{y}$.}
\begin{equation*}
\begin{aligned}
    \zeta &= \mathrm{Re}\{e^{-j\phi}\boldsymbol{y}^H\boldsymbol{a}\boldsymbol{c}^H \boldsymbol{y}\} \\ &= \boldsymbol{y}^H \left( \frac{e^{-j\phi}\boldsymbol{a}\boldsymbol{c}^H + e^{j\phi}\boldsymbol{c}\boldsymbol{a}^H}{2}\right) \boldsymbol{y} \triangleq \boldsymbol{y}^H\boldsymbol{M}\boldsymbol{y} . 
    \end{aligned}
\end{equation*}
Let us assume the Rx has the perfect information about two directions $\boldsymbol{a}$ and $\boldsymbol{c}$. 
In what follows, we obtain the distribution of $\zeta$ and investigate the detection threshold as well as error performance of the coherent receiver.

In Appendix \ref{appendix:ALD_PDF}, it is shown that the distribution of $\zeta$ conditioned on $x_m^{}, m\in\{0,1\}$ follows Asymmetric Laplace Distribution (ALD)  \cite[Chapter~3]{Kotz2001} of which cumulative density function (CDF) and probability density function (PDF) are given by
\begin{subequations}
\begin{align}
\label{eq:CDF_real}
    F(\zeta| x_m^{}) &= \begin{cases} 
      -\frac{\lambda_1(x_m^{})}{\lambda_2(x_m^{}) - \lambda_1(x_m^{})}  e^{-\frac{\zeta}{\lambda_1(x_m^{})}} & \zeta < 0 \\
      1 - \frac{ \lambda_2(x_m^{})}{\lambda_2(x_m^{}) - \lambda_1(x_m^{})} e^{-\frac{\zeta}{\lambda_2(x_m^{})}} & \zeta \geq 0 
  \end{cases} , \\
  \label{eq:PDF_real}
    f(\zeta|x_m^{}) &=  
      \frac{1}{\lambda_2(x_m^{}) - \lambda_1(x_m^{})} 
      \begin{cases} 
      e^{-\frac{\zeta}{\lambda_1(x_m^{})}} & \zeta < 0 \\
      e^{-\frac{\zeta}{\lambda_2(x_m^{})}} & \zeta \geq 0  
      \end{cases} ,
  \end{align}
\end{subequations}
respectively, where $\lambda_{\ell}(x_m^{}), \forall \ell \in \{1,2\}$ two eigenvalues of $\mathcal{M} = \boldsymbol{R}_{\boldsymbol{y}|x_m^{}}\boldsymbol{M}$, and $\boldsymbol{R}_{\boldsymbol{y}|x_m^{}}$ denotes the covariance matrix of received signal $\boldsymbol{y}$ conditioned on the tag signal $x_m^{}$.

\begin{figure}
    \centering
    \includegraphics[width=0.42\textwidth]{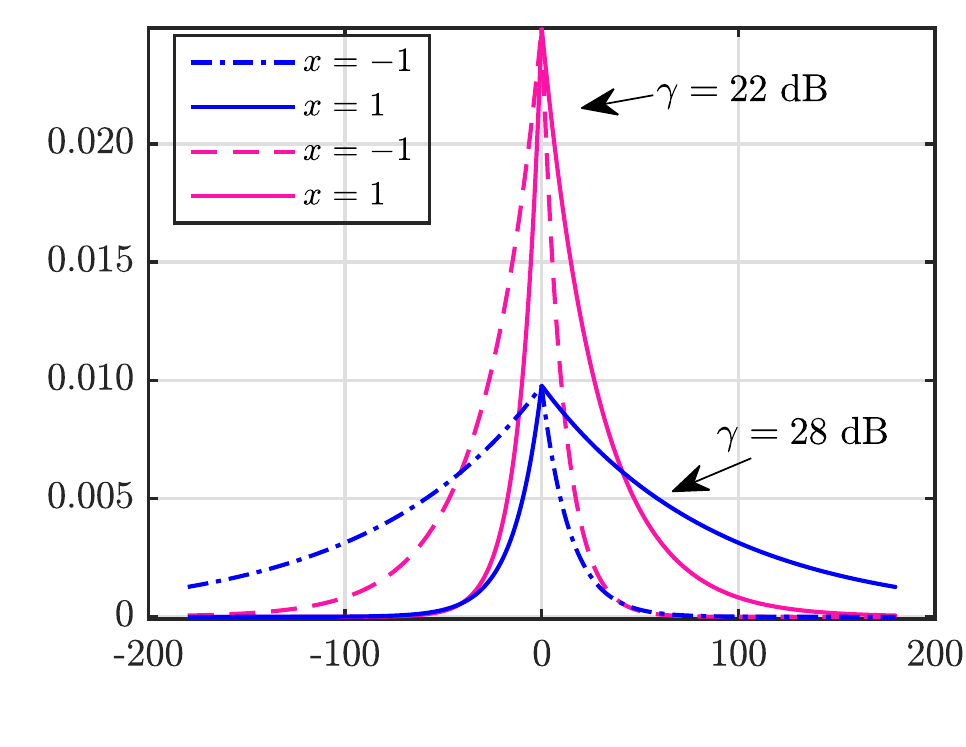}
    \caption{Probability density functions of $\zeta$ conditioned on BPSK-modulated tag signal for different SNR of legacy system $\gamma$}
    \label{fig:PDFofZeta}
\end{figure}

The conditional PDF of the quadratic form written in Eq.~\eqref{eq:PDF_real} indicates that $\zeta$ follows ALD with location parameter 0, scale parameter $- \sqrt{-1/[\lambda_1(x_m)\lambda_2(x_m)]}$ and asymmetry parameter $\sqrt{-\lambda_1(x_m)/\lambda_2(x_m)}$. The expectation and variance of this distribution are $\mathrm{E}\{\zeta|x_m\} = -\lambda_2(x_m)-\lambda_1(x_m)$ and $\mathrm{Var}\{\zeta|x_m\}= \lambda_2^2(x_m) + \lambda_1^2(x_m)$, respectively. Two examples of PDF of two ALD random variables are illustrated in Fig.~\ref{fig:PDFofZeta} for different SNR of legacy system $\gamma$ when the Tag uses BPSK modulation. 

The two eigenvalues of the matrix $\mathcal{M}$ can be more easily written in terms 
\begin{subequations}\label{eq:eigenvalue-parts}
\begin{align}
\varepsilon_m^{} &= \mathrm{Re}\left\{\frac{\gamma}{2}\eta_2^{}\left(  e^{j \phi}\eta_1^{} |x_m^{}|^2 + x_m^* \right)\right\},\\ 
A_m^{}  &= \frac{\gamma}{4}\Big(1 + \Delta  |x_m^{}|^2 +  2\mathrm{Re}\{e^{ j \phi}\eta_1^{} x_m^{}\} + \frac{1}{\gamma} \Big),
\end{align}
\end{subequations}
so that the eigenvalues read as
\begin{equation}
    \lambda_{\ell}(x_m) = \varepsilon_m^{} +(-1)^{\ell} \sqrt{\varepsilon_m^2 +A_m^{}}, \quad\ell \in\{ 1,2\}.
    \label{eq:simplifedEigValue}
\end{equation}
Using the quantities in Eq.~\eqref{eq:eigenvalue-parts}, let us investigate the behaviour of the eigenvalues under practical values of the physical parameters. For a practical AmBC signal when $\eta_2 \neq 0$, $|\eta_1|^2 + |\eta_2|^2 = \Delta \ll 1$ so that $|x_m| \gg |\eta_1|$, which implies $\varepsilon_m \approx \mathrm{Re}\{\frac{\gamma}{2}\eta_2^{}x_m^*\}$. The same line of reasoning leads to $A_m \approx \frac{\gamma}{4}$, which in turn implies $(\varepsilon_m^2 +A_m^{}) \approx A_m$. Then, the eigenvalues can be approximated by
\begin{equation}
\lambda_{\ell}(x_m) \approx \mathrm{Re}\{\frac{\gamma}{2}\eta_2^{}x_m^*\} + (-1)^{\ell}\frac{\sqrt{\gamma}}{2} .
\label{eq:approximateEigValue}
\end{equation}

In Appendix~\ref{appendix:ALD_PDF}, we show that variables $\zeta|x_0$ and $\zeta|x_1$ have opposite expectations, and have similar variances, and enlarging the difference between two expectations improves the demodulation performance. In order to see the conditions on how two expectations can be moved further apart, it is possible to use the approximation of eigenvalues in Eq.~\eqref{eq:approximateEigValue}. Using this approximation, the expectation can be written as $\mathrm{E}\{\zeta|x_m\} = -\mathrm{Re}\{\gamma\eta_2 x_m^*\}$, which implies that the detection performance can be improved by either increasing SNR of the legacy system $\gamma$ or component of backscatter path perpendicular to the direct path $\eta_2$.

The binary tag signal demodulation problem is formed as a binary hypothesis testing, and optimal receivers can be constructed starting from the MAP criterion. 
When the tag signal has equal probability of transmitting $x_0$ or $x_1$, the MAP criterion is equivalent to the maximum likelihood criterion, which compares likelihood probabilities of the testing statistics for a given tag signal value, that is
\begin{equation}
     f(\zeta|x_0) \mathrel{\substack{\mathcal{H}_0\\ \gtrless \\ \mathcal{H}_1}}  f(\zeta|x_1) .
    \label{eq:MLdetection}
\end{equation}
Taking the logarithm of Eq.~\eqref{eq:MLdetection}, we have
\begin{equation*}
    \begin{cases}
    \zeta             \mathrel{\substack{\mathcal{H}_0\\ \gtrless \\ \mathcal{H}_1}} 
    \frac{\lambda_1(x_0) \lambda_1(x_1)}{\lambda_1(x_0) - \lambda_1(x_1)} \ln \frac{\lambda_2(x_0) - \lambda_1(x_0)}{\lambda_2(x_1) - \lambda_1(x_1)} ,  \zeta < 0  
            \nonumber \\
    \zeta \mathrel{\substack{\mathcal{H}_1\\ \gtrless \\ \mathcal{H}_0}} 
    \frac{\lambda_2(x_0) \lambda_2(x_1)}{\lambda_2(x_0) - \lambda_2(x_1)} \ln \frac{\lambda_2(x_0) - \lambda_1(x_0)}{\lambda_2(x_1) - \lambda_1(x_1)} ,  \zeta \geq 0
    \end{cases} .
\end{equation*}
This test yields two thresholds with different signs as
\begin{equation*}
 \begin{aligned}
     T_1 &= \frac{\lambda_1(x_0) \lambda_1(x_1)}{\lambda_1(x_0) - \lambda_1(x_1)} \ln \frac{\lambda_2(x_0) - \lambda_1(x_0)}{\lambda_2(x_1) - \lambda_1(x_1)} < 0 ,\\
     T_2 &= \frac{\lambda_2(x_0) \lambda_2(x_1)}{\lambda_2(x_0) - \lambda_2(x_1)} \ln \frac{\lambda_2(x_0) - \lambda_1(x_0)}{\lambda_2(x_1) - \lambda_1(x_1)} \geq 0.
 \end{aligned}
\end{equation*}
It can be observed that the thresholds depend on the ratio of $\lambda_1(x_0) - \lambda_1(x_1)$ and $\lambda_2(x_0) - \lambda_2(x_1)$. Then, Eq.~\eqref{eq:approximateEigValue}
implies that 
\begin{equation*}
    \lambda_{\ell}(x_0)  - \lambda_{\ell}(x_1)\approx \frac{\gamma}{2}\eta_2^{}\mathrm{Re}\{x_0^* - x_1^*\},
\end{equation*}
so the sign of eigenvalue differences can be judged by looking at the difference $x_0^* - x_1^*$. For the modulation schemes that we consider in this paper\footnote{For BPSK modulation, $x_0 = -1$ and $x_1 = +1$; for OOK modulation $x_0 = 0$ and $x_1 = +1$}, we always have $x_0 < x_1$, and the domains of the thresholds imply that the error probability  should be calculated under two different cases:
\begin{enumerate}

    \item $\lambda_1(x_0) - \lambda_1(x_1) \leq \lambda_2(x_0) - \lambda_2(x_1) < 0$ :
    \begin{equation*}
    \begin{aligned}
      &\qquad   T_2 < 0 \implies \left(\zeta 
        \mathrel{\substack{\mathcal{H}_1\\ \gtrless \\ \mathcal{H}_0}}
         T_1    \right) \implies \\
       &\begin{aligned} p_e^{} = & \frac{1}{2} \left [\int_{-\infty}^{T_1} f(\zeta|x_1) d\zeta  +\int_{T_1}^{\infty} f(\zeta|x_0) d\zeta \right] \end{aligned},
        \end{aligned}
    \end{equation*}
    
    \item  $\lambda_2(x_0) - \lambda_2(x_1) < \lambda_1(x_0) - \lambda_1(x_1) < 0$ :
    \begin{equation*}
    \begin{aligned}
         & \qquad T_1 \ge 0 \implies \left( \zeta 
        \mathrel{\substack{\mathcal{H}_1\\ \gtrless \\ \mathcal{H}_0}}
        T_2 \right) \implies \\
        &\begin{aligned}p_e^{} = & \frac{1}{2} \left [\int_{-\infty}^{T_2} f(\zeta|x_1) d\zeta  +\int_{T_2}^{\infty} f(\zeta|x_0) d\zeta \right] \end{aligned} .
        \end{aligned}
    \end{equation*}
\end{enumerate}
Another situation occurs when $\eta_2 = 0$, which leads $\varepsilon_m = 0$, $A_m \approx \frac{\sqrt{\gamma}}{2}|1 + \eta_1 x_m|$. In this case, the probability of error should be calculated using
\begin{enumerate}
\setcounter{enumi}{2}

\item $\lambda_1(x_0) - \lambda_1(x_1) > 0 > \lambda_2(x_0) - \lambda_2(x_1)$ :
    \begin{equation*}
        \begin{aligned}
        &\qquad T_1< 0 ~\text{and} ~T_2 \geq 0  \implies\\
      &\begin{aligned}   
      p_e  =  \frac{1}{2}\Bigg[&\int_{-\infty}^{T_1} f(\zeta|x_0)d\zeta\\& + \int_{T_1}^{T_2} f(\zeta|x_1) d\zeta  + \int_{T_2}^{\infty} f(\zeta|x_0) d\zeta\Bigg].      \end{aligned}
        \end{aligned}
    \end{equation*}
\end{enumerate}
This condition implies that the BD is on the direction $\boldsymbol{a}$, and the receiver cannot discriminate it from the ambient signal. When the number of receiver antennas $N_r$ is large, this might only happen when the tag is at the transmitter. As $N_r$ gets smaller, this condition is observed when the tag is on the line connecting the Tx to Rx. 

\begin{figure}[t!]
    \centering
    \includegraphics[width=0.465\textwidth]{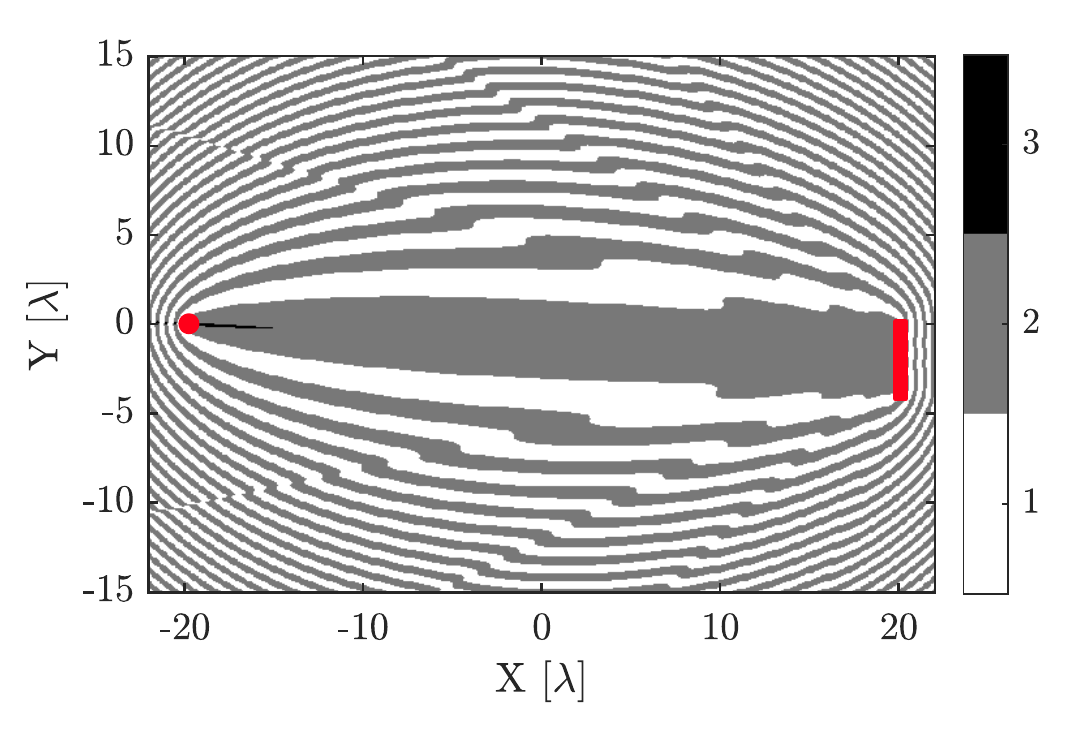}
    \caption{Variation of the error probability integral boundary conditions as a function of tag position for direct path length $d_{01} = 40 \lambda$ and number of antennas $N_r = 8$. The Tx antenna and the Rx antenna array are marked as red circle and red bar, respectively.  }
     \label{fig:Pe_calculation_case}
\end{figure}

The decision thresholds $T_1$ and $T_2$ are functions of the components of backscatter path $\eta_1$ and $\eta_2$. Thus the decision threshold varies with tag locations, and in turn, error probability integral boundaries enumerated above change. The variation of the effective condition as a function of location of the tag is visualized in Fig.~\ref{fig:Pe_calculation_case}. Although the first two conditions vary similar to ellipses shown in Fig.~\ref{fig:phase_location}, the third condition is only observed when the tag is on the line between the Tx and the Rx. 

In practical implementations, the receiver calculates the detection thresholds, $T_1$ and $T_2$, and four eigenvalues using the preamble measurements $\boldsymbol{Y}_{t+}$,  $\boldsymbol{Y}_{t-}$, and the direction estimates $\hat{\boldsymbol{a}}$ and $\hat{\boldsymbol{c}}$.

\subsection{Non-coherent receiver}
Another typical method to detect the tag signal with unknown parameters is non-coherent receiver, which require statistical information on the unknown parameters. The non-coherent receivers marginalize (or average) over unknown parameters which, in our case, are ambient signal $s$ and phase offset $\phi$. Recalling that we have obtained the effective backscatter signal through two-stage beamforming given in Eq.~\eqref{eq:relevantSignal}, the likelihood function of $u$ given $x_m$ can be written as
\begin{equation*}
\begin{aligned}
   p(u|x_m) &=
  \iint f(u|s,x_m,\phi)f(s)f(\phi)ds d\phi \\
  &= \frac{1}{\pi\left(\gamma|\eta_2|^2|x_m|^2+1\right)}   \exp\left\{ -\frac{|u|^2}{\gamma|\eta_2|^2|x_m|^2+1}\right\} . 
 \end{aligned}
\end{equation*}
Substituting it into the maximum likelihood criterion expressed in Eq.~\eqref{eq:ML_criterion} and taking logarithm on both sides yields
\begin{equation*}
 \begin{aligned}
   \frac{|u|^2}{\gamma|\eta_2|^2|x_0|^2+1} + & \ln\left(\gamma|\eta_2|^2|x_0|^2+1\right)
         \mathrel{\substack{\mathcal{H}_0\\ \lessgtr \\ \mathcal{H}_1}} \\ &
        \frac{|u|^2}{\gamma|\eta_2|^2|x_1|^2+1} +\ln\left(\gamma|\eta_2|^2|x_1|^2+1\right) \\
   \implies  \quad  |u|^2   \mathrel{\substack{\mathcal{H}_0\\ \lessgtr \\ \mathcal{H}_1}}  ~ T_h & ,
 \end{aligned} 
\end{equation*}
which shows that the sufficient statistic of the non-coherent receiver is $|u|^2$. In other words, the optimum non-coherent receiver is the energy detector when the statistics of $s$ and $\phi$ follows the assumed distributions.  
The threshold of the test $T_h$ is expressed as
\begin{equation*}
\begin{aligned}
 T_h  = ~ &\frac{\left(\gamma|\eta_2|^2|x_0|^2+1 \right) \left(\gamma|\eta_2|^2|x_1|^2+1 \right)}{\gamma|\eta_2|^2 (|x_1|^2-|x_0|^2)}\cdot \\ &\qquad\ln\left( \frac{\gamma|\eta_2|^2|x_1|^2+1}{\gamma|\eta_2|^2|x_0|^2+1}\right) .
  \end{aligned}
\end{equation*}
It can be seen that $T_h$ goes to infinity when tag is BPSK-modulated, which implies that the non-coherent receiver is useful only when the tag changes the amplitude of $x$, e.g., by using OOK. In this case, the measurement is tested for presence of the tag signal against absence of it, and the threshold is
\begin{equation*}
    T_h = \left(1+\frac{1}{\gamma|\eta_2|^2 } \right) \ln\left( \gamma|\eta_2|^2 + 1\right) .
\end{equation*}

The non-coherent receiver derivations given above can be used for receiver performance analysis for OOK-modulated tag signal. Under $\mathcal{H}_0$, the effective signal $u = \boldsymbol{c}^H\boldsymbol{\omega}$, and it is easy to see that $2|u|^2$ follows the chi-square distribution with 2 degrees of freedom\footnote{The coefficient 2 in front of the testing statistic is due to the fact that the random noise has common variance equals to  $1/2$ per real and imaginary components.}, i.e., $2|u|^2 \sim \chi_2^2$. Based on this statistic, the probability of false alarm is given by~\cite[Section~2.3]{Proakis2001}
\begin{equation*}
    P_f = \mathbb{P}\left\{|u|^2>T_h |\mathcal{H}_0\right\} = 1- \Bar{\Gamma}(1,T_h) ,
\end{equation*}
where $\Bar{\Gamma}(s,x)$ is the lower incomplete Gamma function
\begin{equation*}
   \Bar{\Gamma}(s,x) =  \int_{0}^{x} t^{s-1} e^{-t} dt.
\end{equation*}
Under $\mathcal{H}_1$, Eq.~\eqref{eq:relevantSignal} is written as $u = \sqrt{\gamma}\eta_2 s + \boldsymbol{c}^H\boldsymbol{\omega}$ which follows $\mathcal{CN}(0, \gamma\eta_2^2 +1)$. 
Then, $\frac{2}{\gamma\eta_2^2 +1}|u|^2$ follows the chi square distribution with 2 degrees of freedom. The probability of miss detection for this statistic is 
\begin{equation*}
    P_M = \mathbb{P}\left\{|u|^2\leq T_h |\mathcal{H}_1\right\} = \Bar{\Gamma}\left(1, \frac{T_h}{\gamma\eta_2^2 + 1}\right).
\end{equation*}
Hence, the error probability is given by
\begin{equation}
\begin{aligned}
    p_e =\frac{1}{2}\left[1 +\Bar{\Gamma}\left(1, \frac{T_h}{\gamma\eta_2^2 + 1}\right) -  \Bar{\Gamma}(1,T_h)\right] .
\end{aligned}
\label{eq:error_probability_noncoherent}
\end{equation}

\begin{figure}[t!]
    \centering
    \includegraphics[width=0.465\textwidth]{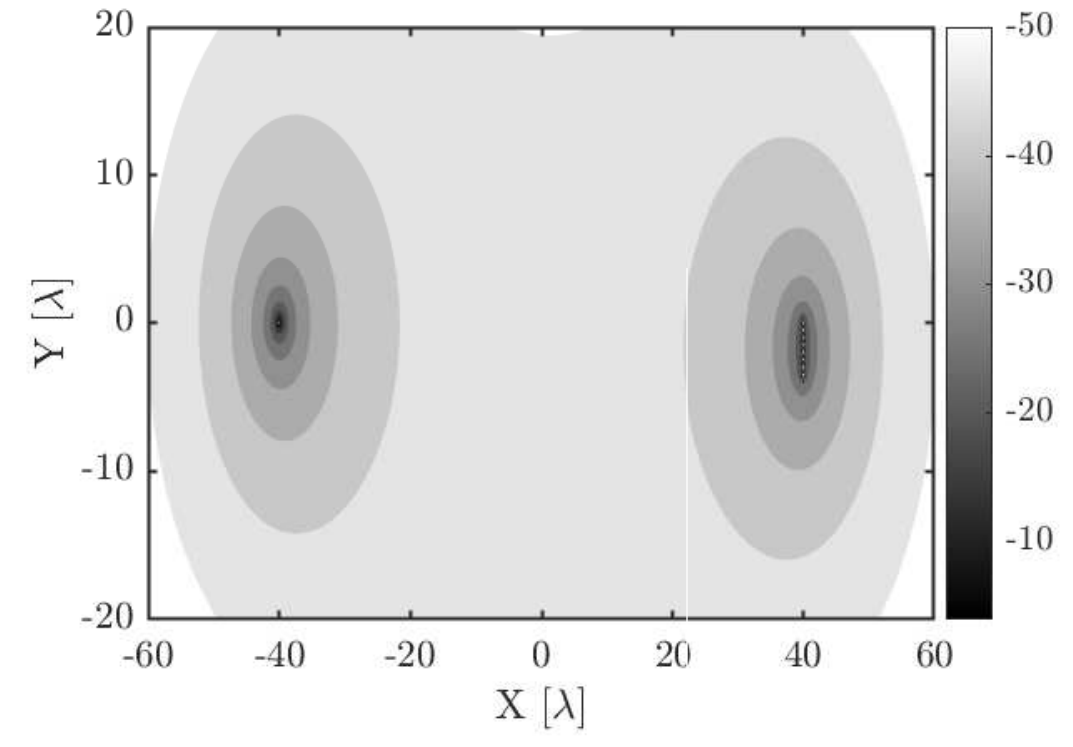}
    \caption{Variation of power difference between two paths $\Delta$ as a function of tag position for direct path length $d_{01} = 80 \lambda$ and number of antennas $N_r = 8$  }
     \label{fig:powerDifference_location}
\end{figure}

The error probability in Eq.~\eqref{eq:error_probability_noncoherent} is defined by  $\gamma|\eta_2|^2$ which represents the effective SNR of the backscatter signal. Thus, detection performance can be improved by  either increasing the legacy system SNR $\gamma$ or component of backscatter path perpendicular to the direct path $\eta_2$. In other words, when $\eta_2 = 0$, i.e., the direct path and the backscatter path are along the same direction, the non-coherent receiver has the worst performance. This fact suggests that the tag should not be placed on the direct path also for non-coherent receiver.

\section{Simulation results}\label{sec:results}

\begin{figure*}
\centering
	\centering
	\setlength{\tabcolsep}{0pt}
	\begin{tabular}{ccc}
		\subfloat[]{\includegraphics[width=0.32\textwidth]{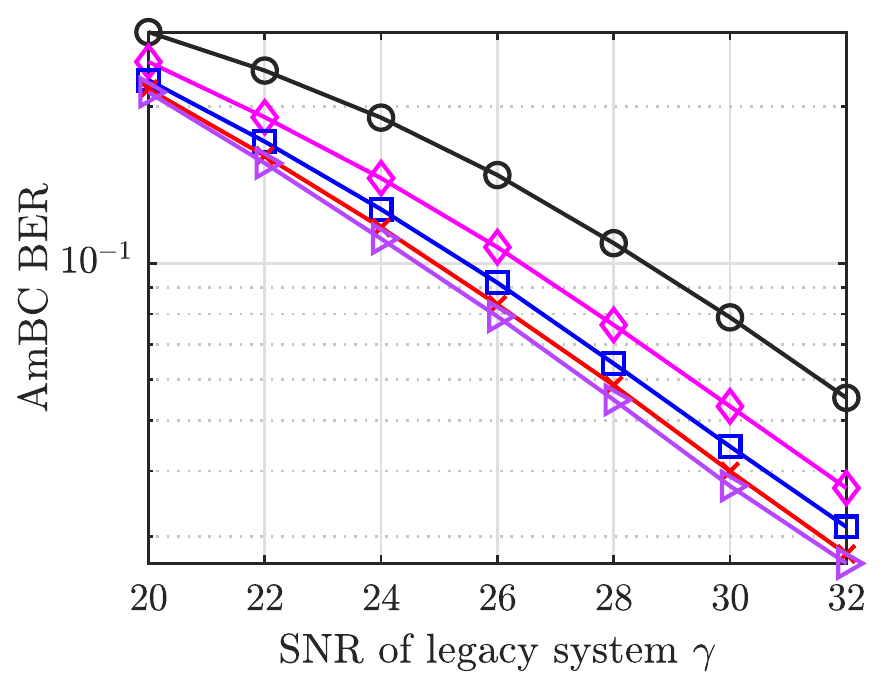}
			 \label{fig:ber_Nr}} &
		\subfloat[]{\includegraphics[width=0.32\textwidth]{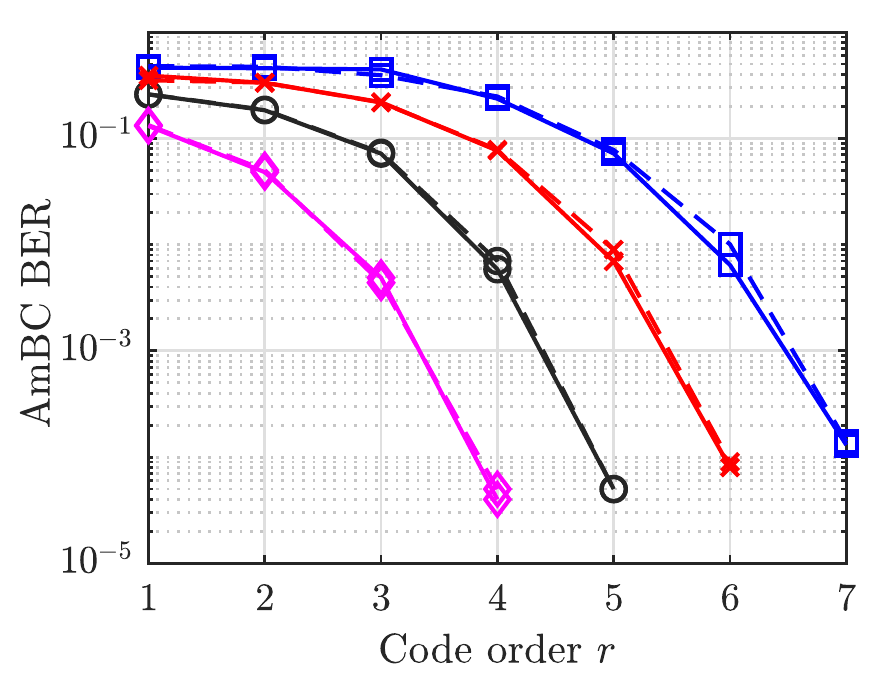} 
			\label{fig:ber_r}} &
		\subfloat[]{\includegraphics[width=0.32\textwidth]{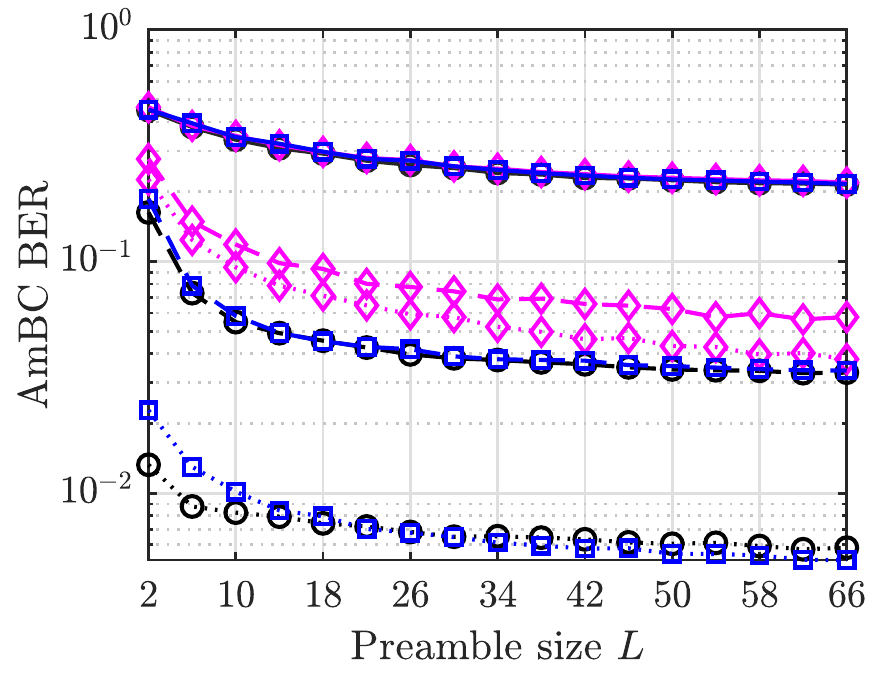} 
			\label{fig:ber_L}}
	\end{tabular}
\caption{In (a), variation of AmBC Bit-Error-Rate (BER) as a function of the legacy system SNR $\gamma$ for preamble size $L=64$ and power difference between two paths $\Delta \approx -31$dB with different markers representing the number of antennas $N_r$: $N_r = 2$ (\protect\blackcircle), 4 (\protect\magentadiamond), 6 (\protect\bluesquare), 8 (\protect\redcross) and 10 (\protect\purpletriangle). 
In (b), variation of AmBC BER as a function of the code order $r$ for $L=64$ and $\gamma = 28$ dB. The solid lines represent Hadamard coding while dashed lines represent Simplex coding with different markers for distance between tag and the receiver $d_{11}$: $d_{11} = 2\lambda$ (\protect\magentadiamond), $4\lambda$ (\protect\blackcircle), $6\lambda$ (\protect\redcross) and $8\lambda$ (\protect\bluesquare). 
In (c), variation of AmBC BER as a function of $L$ for $\gamma = 28$ dB with different line types for different $\Delta$: $ \approx$-39 dB (solid), $ \approx$-28 dB (dashed) and $ \approx$-18 dB (dot) and different markers for machine learning algorithms: linear discriminant analysis (\protect\magentadiamond), logistic regression (\protect\blackcircle) and soft margin support vector machine (\protect\bluesquare).}
\label{fig:ML-assistedVsParameters}
\end{figure*}

In this section, simulation results are provided to validate the performance of the proposed receiver. The distances are wavelength-scaled in order to make the result carrier-frequency-independent.  All the results are obtained by averaging over $10^6$ Monte Carlo realizations. In the following,
we first evaluate the performance of the proposed ML-assisted receiver with different parameters to show their impact. Then, we compare the performance of ML-assisted method with the traditional receivers. Finally, we provide the variation of the performance with the tag location to show the expected coverage area for a single-tag deployment. 

\subsection{Numerical evaluation}

We consider a linear antenna array at the Rx with half-wavelength  $\lambda/2$ antenna separation. The Tx and Rx are separated by the distance $d_{01} = 80\lambda$ as shown in Fig. \ref{fig:systemModel}. 

The variation of power difference $\Delta$ between two paths as a function of tag location for $N_r = 8$ is plotted in Fig. \ref{fig:powerDifference_location}. It can be seen that the backscatter path undergoes a tremendous power loss, nearly -50 dB, when the tag is far away from both the Tx and Rx. Even with a short distance $d_{11} \approx 2\lambda$, $\Delta$ already reaches to -30 dB. This result shows the importance of improving the effective SNR of the backscatter path.

Hereafter, the incident angle between the backscatter path and the antenna array is fixed to $\pi/4$ radians but with varying distance $d_{11}$, i.e. the tag location is $\boldsymbol{p}=[10\lambda-d_{11}/\sqrt{2}, d_{11}/\sqrt{2}]$. 
BER-performances of the proposed ML-assisted method for BPSK-modulated tag signal with different parameters are shown in Fig. \ref{fig:ML-assistedVsParameters}. In Fig. \ref{fig:ber_Nr}, BERs of ML-assisted method without using coding and  preamble size is set to $L = 64$ with different number of antennas $N_r$ are compared. The result clearly shows that the improvement of increasing number of receiver antennas is diminishing as $N_r$ becomes larger. This is consistent with the error probability of spatial diversity~\cite[Section~3.3]{Tse2005}. Considering this result, in the following, we fix $N_r = 8$.

\begin{figure}[t!]
    \centering
    \includegraphics[width = 0.42\textwidth]{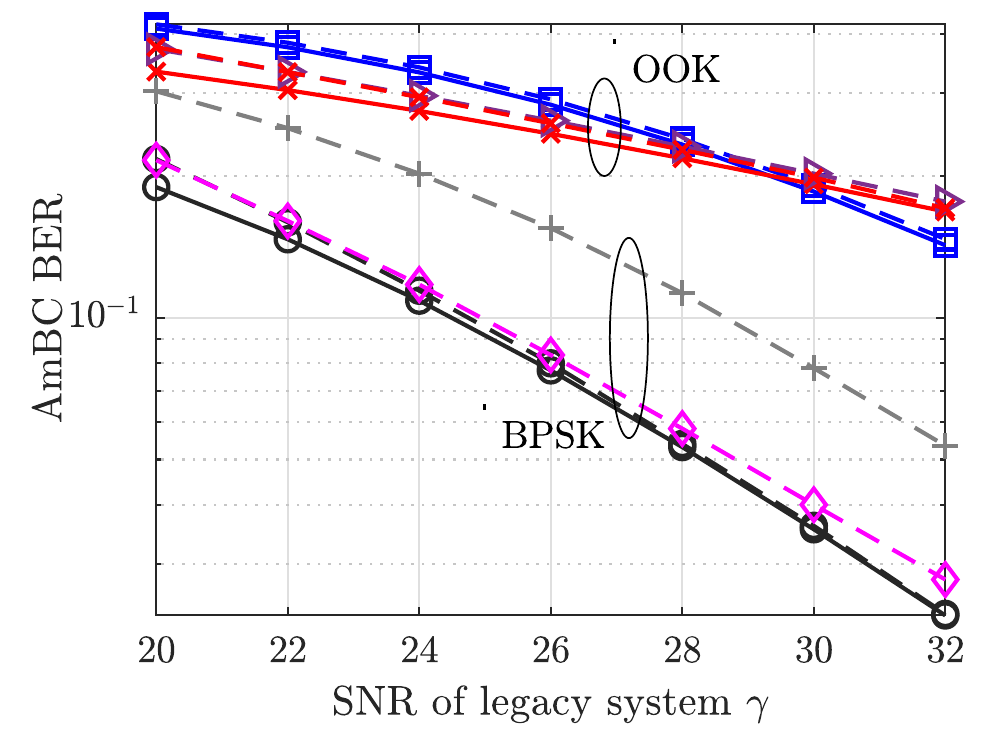}
    \caption{Variation of AmBC Bit-Error-Rate (BER) as a function of the SNR of legacy system for $\Delta \approx -31 dB$ with different markers for different receivers:  coherent receiver for OOK modulation (\protect\redcross),  non-coherent receiver for OOK modulation (\protect\bluesquare), coherent receiver for BPSK modulation(\protect\blackcircle), ML-assisted receiver for OOK modulation (\protect\purpletriangle), ML-assisted receiver for BPSK modulation (\protect\magentadiamond) and ignoring phase offset (\protect\grayplus). The solid lines illustrate the theoretical error probabilities while the dashed lines represent the simulated error probabilities.   }
    \label{fig:co_nonco}
\end{figure}

The effect of coding on BER as a function of code order $r$ is shown in Fig.~\ref{fig:ber_r} for the SNR of legacy system $\gamma = 28$ dB and preamble size $L = 64$. It can be seen that using longer codewords improves the BER-performance as longer codewords can correct more errors. Furthermore, the Hadamard code and the Simplex code obtain a similar performance under the same conditiona. Since Simplex code has one dimension less than Hadamard code, it has a higher energy per bit when $r$ is small. Hence, the tag can adopt Simplex coding to preserve energy, and for the remaining results, we use Simplex coding.

The impact of preamble size $L$ on BER-performance is illustrated in Fig.~\ref{fig:ber_L}. Solid, dashed and dot lines represent the effective SNR of -11, 0, and 10 dB, respectively.  The results indicate that increasing the training set length improves the BER since longer preambles provide more samples for classifiers as well as better estimates of directions $\boldsymbol{a}$ and $\boldsymbol{c}$. However, the improvement becomes minor when $L > 34$.  In addition, a longer training sequence consumes more energy of the tag and adds computational complexity at the Rx. Hence, it is reasonable to choose $L = 34$. The performance difference between LDA and LR becomes larger as effective SNR increases. This is because LDA works under the assumption of Gaussian distribution which is not true in our case as has been elaborated on in Sec.\ref{sec:conventional_receivers}. The soft margin SVM  and LR have similar performance, but the soft margin SVM slightly outperforms LR when effective SNR is 10 dB and $L>=34$ since LR is more susceptible to outliers. As soft margin SVM requires tuning of a hyperparameter, LR is preferable among studied ML algorithms.

\subsection{Performance comparison}

In Fig. \ref{fig:co_nonco}, performances of the proposed ML-assisted receiver, the non-coherent receiver, and the coherent receiver as a function of legacy system SNR $\gamma$ are compared for both OOK and BPSK modulations where coding is not used.  The tag is $2\lambda$ away from the Rx such that the power difference $\Delta\approx -31$ dB. Solid lines are error probabilities of the conventional receivers analyzed in Section \ref{sec:conventional_receivers}, while dashed lines with the same markers are their corresponding numerical results.  It is observed that BPSK modulation has more than 6 dB gain compared with the OOK modulation. For OOK modulation, the non-coherent receiver does not work in low SNR region and it slightly outperforms the coherent receiver in high SNR region. This is because the coherent receiver coarsely estimates the ambient signal whereas the non-coherent receiver considers its signal space. For both modulations, the proposed ML-assisted receiver performance is close to the performance of the coherent receiver with known phase offset which provides a lower bound of the error probability. As discussed earlier, ignoring the phase offset loses 3 dB gain from the coherent receiver as shown by the dashed line with marker (\protect\grayplus). The result shows that it is necessary to take into account the phase offset, and the proposed receiver sufficiently mitigates its adverse impact.

\begin{figure}[t!]
    \centering
    \includegraphics[width=0.465\textwidth]{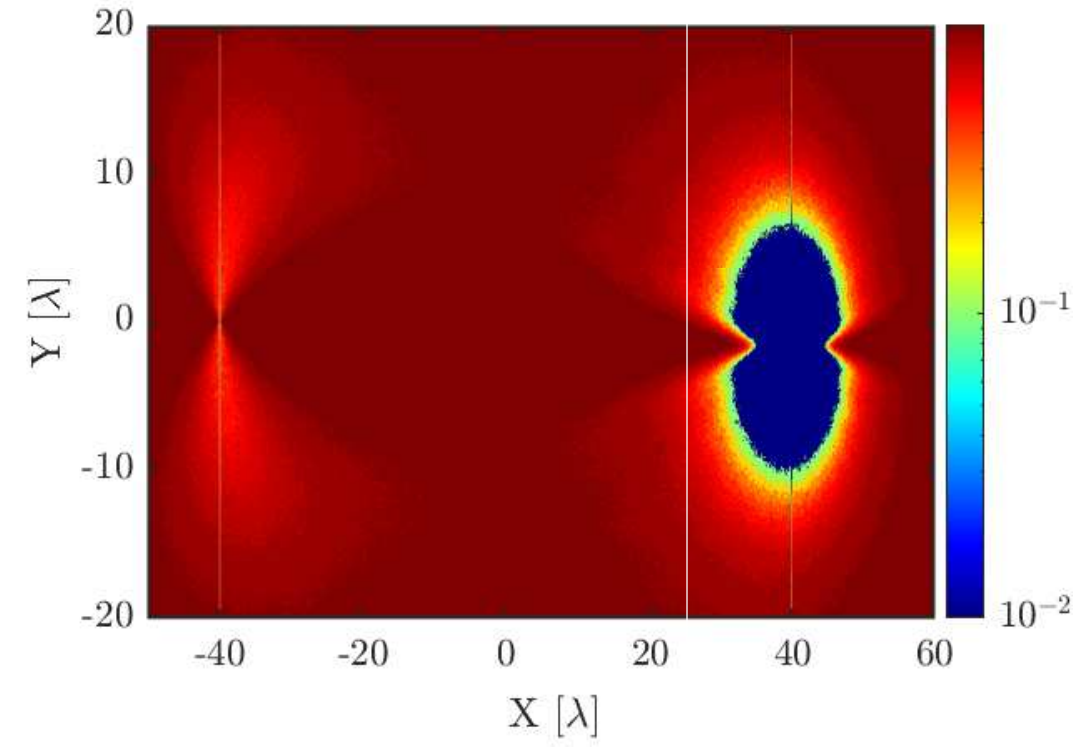}
    \caption{Variation of AmBC symbol error rate in log scale as a function of tag location for $d_{01} = 80 \lambda$, code order $r=3$, and legacy system SNR of $28$ dB}
     \label{fig:ber_location}
\end{figure}

\subsection{Coverage area}
In Fig. \ref{fig:ber_location}, variation of the proposed receiver symbol error rate (SER) performance in log scale for code order $r = 3$ and legacy system SNR $\gamma=28$ dB is shown to gain a perspective of coverage area of a single-tag deployment. The tag is placed within a $(110\lambda \times 40\lambda)$ area in which the Tx and Rx are placed at $\boldsymbol{p}_t = [-40\lambda,0]$ and $\boldsymbol{p}_r = [40\lambda,0]$, respectively. The SER increases as both the distance between tag and Tx, and distance between tag and Rx increase, which is coherent with the result in Fig. \ref{fig:powerDifference_location}. As can be seen from the figure, there is a null beam, i.e., worse SER, on the line between Tx and Rx.  In this area, $\boldsymbol{a}$ and $\boldsymbol{h}$ are inseparable, which gives rise to the fact that the backscatter path is also canceled while nullifying the direct path inference. The shape of the coverage area close to the Rx is caused by the symmetric linear antenna array. The result shows the coverage area is in the close vicinity of the Tx and a large region ($\sim 20\lambda$) around the Rx.

\section{Conclusion}\label{sec:conclusion}
In this paper, the coherent reception of the BPSK-modulated tag signal in AmBC systems is studied. The testing statistic is derived from MAP criterion, which does not depend on the prior information of ambient signal. A coherent receiver that utilizes multiple antennas to mitigate the strong DPI and the rapidly varying ambient signal, and uses logistic regression algorithm to learn the pattern of phase offset caused by excess length of the backscatter path is presented. The closed-form error probability of the coherent receiver with perfect phase offset and non-receiver for OOK modulation are derived in order to judge the achieved improvement of the proposed method. The designed receiver achieves the same BER-performance with 1-dB more SINR compared with the ideal coherent receiver, and outperforms the non-coherent receiver. The work in this paper suggests that a multi-antenna receiver can effectively mitigate the direct-path interference, and enable coherent demodulation of the tag signal, which has not been reported in the literature. The receiver performance is defined by legacy system SNR and tag location. The spatial  performance variation shows that the coverage area of a single tag deployment is in the close vicinity of the transmitter and within 20 wavelength around the receiver. Consequently, a successful AmBC network deployment can be achieved by placing the receiver in a suitable location with respect to the location of the tags, and the distance between them can be further increased by using the presented coherent receiver.

\appendices
\section{Proof of Proposition I} \label{appendix:testing_statistic}
The demodulation of BPSK-modulated tag signal is formed by a binary hypothesis testing, written as $\mathcal{H}_m: x_m$ is transmitted, $m \in\{0, 1\}$. When the effective tag signal $u[i]$ is obtained, the binary hypothesis test of the optimum receiver is built upon the MAP criterion
\begin{equation*}
    p(x_0|u[i]) \mathrel{\substack{\mathcal{H}_0\\ \gtrless \\ \mathcal{H}_1}} p(x_1|u[i]), 
\end{equation*}
where $p(x_m|u[i])$ is the posterior probability of $x_m$ given the effective tag signal. According to Bayes' theorem, the MAP criterion is written as
\begin{equation}
\begin{aligned}
    \frac{p(x_0) p(u[i]|x_0)  }{p(u[i])}  &\mathrel{\substack{\mathcal{H}_0\\ \gtrless \\ \mathcal{H}_1}}  \frac{p(x_1) p(u[i]|x_1)}{p(u[i])}
    \\ p(u[i]|x_0) &\mathrel{\substack{\mathcal{H}_0\\ \gtrless \\ \mathcal{H}_1}}  p(u[i]|x_1) ,
    \label{eq:ML_criterion}
    \end{aligned}
\end{equation}
where $p(u[i])$ is irrelevant for making a decision of $x_m$, and $x_m$ have equal probability such that the MAP criterion becomes maximum likelihood criterion. The likelihood function $p(u[i]|x_m)$ contains a hidden variable, i.e., the ambient signal $s$. If the statistical information of the unknown $s[i]$ is provided, we can calculate $p(u[i]|x_m)$ by integrating the joint distribution over the ambient signal space. In our case, the ambient signal can be partially estimated as
\begin{equation*}
    \hat{s}[i] = \boldsymbol{a}^H \boldsymbol{y}[i] .
\end{equation*}
We can now rewrite $p(u[i]|x_m)$ as
\begin{equation*}
\begin{aligned}
   & ~p(u[i]|x_m) = \int p(u[i],s|x_m) ds  \\
   =  &\int p(u[i]|s, x_m) p(s) ds = p(u[i]|s=\hat{s}[i], x_m) . 
\end{aligned}
\end{equation*}
The last equality holds since the ambient signal is coarsely estimated such that  $p(s[i]=\hat{s}[i]) = 1$. Substituting it into Eq.~\eqref{eq:ML_criterion} yields
\begin{equation*}
\begin{aligned}
    p(u[i]|s=\hat{s}[i], x_0) &\mathrel{\substack{\mathcal{H}_0\\ \gtrless \\ \mathcal{H}_1}} p(u[i]|s=\hat{s}[i], x_1) 
    \\
    -|u[i] - \sqrt{\gamma} e^{j\phi} \eta_2 \hat{s}[i]x_0|^2 &\mathrel{\substack{\mathcal{H}_0\\ \gtrless \\ \mathcal{H}_1}} - |u[i] - \sqrt{\gamma}  e^{j\phi} \eta_2 \hat{s}[i] x_1|^2  \\
    \mathrm{Re}\{e^{-j\phi}  \boldsymbol{y}[i]^H\boldsymbol{a}\boldsymbol{c}^H \boldsymbol{y}[i]x_0\} 
    &\mathrel{\substack{\mathcal{H}_0\\ \gtrless \\ \mathcal{H}_1}} \mathrm{Re}\{ e^{-j\phi}  \boldsymbol{y}[i]^H\boldsymbol{a}\boldsymbol{c}^H \boldsymbol{y}[i]x_1\} ,
\end{aligned}
\end{equation*}
wherethe common real coefficient $2\sqrt{\gamma} \eta_2$ is cancelled out.  

\section{Distribution of testing statistic}
\label{appendix:ALD_PDF}
The received signal $\boldsymbol{y}$ given $x_m$ can be written as
\begin{equation*}
    \boldsymbol{y}|x_m^{} = \boldsymbol{R}_{\boldsymbol{y}|x_m^{}}^{\frac{1}{2}} \boldsymbol{\nu} , 
\end{equation*}
where $\boldsymbol{\nu} \sim \mathcal{CN}\{\boldsymbol{0},\boldsymbol{I}\}$ is a standard circularly symmeteric complex Gaussian random vector. The covariance matrix when $x=x_m^{}$ is 
\begin{equation*}
\begin{aligned}
    \boldsymbol{R}_{\boldsymbol{y}|x_m^{}} &= E \left\{ \boldsymbol{y}\boldsymbol{y}^H |x_m^{} \right\} \nonumber \\
    &= \gamma\cdot \big[ (1\! + |\eta_1^{}|^2 |x_m^{}|^2\!+ e^{j \phi}\eta_1^{} x_m^{} \!+ e^{-j \phi} \eta_1^* x_m^*\!)\boldsymbol{a} \boldsymbol{a}^H 
    \nonumber \\
    &~~~  + ( \eta_1^{} \eta_2^* |x_m^{}|^2 \!+ e^{-j \phi} \eta_2^* x_m^*\!) \boldsymbol{a}\boldsymbol{c}^H
    +  |\eta_2^{}|^2 |x_m^{}|^2 \boldsymbol{c} \boldsymbol{c}^H \nonumber \\
    & ~~~ + (  \eta_1^* \eta_2^{} |x_m^{}|^2 \!+ e^{j \phi} \eta_2^{} x_m^{}\!)\boldsymbol{c}\boldsymbol{a}^H \big] + \boldsymbol{I} ,
    \end{aligned}
\end{equation*}
which is a full rank matrix.
Then, $\zeta$ conditioned on $x=x_m$ can be rewritten as
\begin{equation}
\begin{aligned}
    \zeta|x_m^{} = \boldsymbol{\nu}^H \boldsymbol{R}_{\boldsymbol{y}|x_m^{}}^{1/2}
   \boldsymbol{M} \boldsymbol{R}_{\boldsymbol{y}|x_m^{}}^{1/2} \boldsymbol{\nu} .
   \label{eq:indefiniteGQF}
  \end{aligned}
\end{equation}
Using the result from Al-Naffouri \emph{et. al}~\cite{Al-Naffouri2016}, it is known that the distribution of the quadratic form over Gaussian random variables depends on eigenvalues of $ \boldsymbol{R}_{\boldsymbol{y}|x_m^{}}^{1/2}
   \boldsymbol{M} \boldsymbol{R}_{\boldsymbol{y}|x_m^{}}^{1/2}$. Now, since the non-zero eigenvalues of matrices $\boldsymbol{AB}$ and $\boldsymbol{BA}$ are the same~\cite[Theorem~1.3.22]{Horn}, the non-zero eigenvalues of $ \boldsymbol{R}_{\boldsymbol{y}|x_m^{}}^{1/2}
   \boldsymbol{M} \boldsymbol{R}_{\boldsymbol{y}|x_m^{}}^{1/2}$ can be calculated from matrix $\mathcal{M}$ defined as
\begin{equation}
    \mathcal{M} = \boldsymbol{R}_{\boldsymbol{y}|x_m^{}} \boldsymbol{M}\!  = e_1^{} \boldsymbol{aa}^H\! +e_1^* \boldsymbol{cc}^H \!+ e_2^{} \boldsymbol{ac}^H\!+ e_3^{} \boldsymbol{ca}^H ,
    \label{eq:H}
\end{equation}
where
\begin{equation*}
\begin{aligned}
    e_1^{} &= \frac{\gamma}{2}\left( e^{j\phi} \eta_1^{}\eta_2^* |x_m^{}|^2 +  \eta_2^* x_m^* \right) ,  \\
      e_2^{} &=\!\frac{\gamma}{2}e^{-j\phi}\! \left(1 \!+  |\eta_1^{}|^2 |x_m^{}|^2\!  + e^{j \phi}\eta_1^{} x_m^{}\! + e^{-j \phi} \eta_1^* x_m^*\! +  \frac{1}{\gamma}\!\right) ,  \\
       e_3^{} &= \frac{\gamma}{2}e^{j\phi} \left(  |\eta_2^{}|^2 |x_m^{}|^2 + \frac{1}{\gamma}\right) .
       \end{aligned}
\end{equation*}

Next, let us look at the rank of matrix $\mathcal{M}$. It is easy to see the eigenvalue decomposition of matrix $\boldsymbol{M}$ is 
\begin{equation*}
\begin{aligned}
  \boldsymbol{M} = \begin{bmatrix} \boldsymbol{u}_1^{} \quad \boldsymbol{u}_2^{} \end{bmatrix} \begin{bmatrix} 0.5 &0  \\ 0 &-0.5 \end{bmatrix}\begin{bmatrix} \boldsymbol{u}_1^{} \quad \boldsymbol{u}_2^{} \end{bmatrix}^H  ,
\end{aligned}
\end{equation*}
where
\begin{equation*}
\boldsymbol{u}_1^{} = (\boldsymbol{a} + e^{j\phi}\boldsymbol{c})/\sqrt{2}, \text{and} ~~ \boldsymbol{u}_2^{} = (- e^{-j\phi} \boldsymbol{a} + \boldsymbol{c})/\sqrt{2} .
\end{equation*}
Therefore, $\boldsymbol{M}$ is a rank-2 matrix. 
According to the Sylvester Inequality~\cite{Horn}, the rank of matrix $\mathcal{M}$ holds
\begin{equation*}
\begin{aligned}
    \rank(\boldsymbol{R}_{\boldsymbol{y}|x_m^{}}) &+ \rank(\boldsymbol{M}) - N_r = 2 \leq \rank(\mathcal{M}) \nonumber \\
    &\leq \min\{\rank(\boldsymbol{R}_{\boldsymbol{y}|x_m^{}}), \rank(\boldsymbol{M}) \}= 2.
    \end{aligned}
\end{equation*}
Thus, matrix $\mathcal{M}$ is also a rank-2 matrix.

The only remaining task is to obtain the eigenvalues of matrix $\mathcal{M}$. For this purpose, theorem~\cite[Theorem~1.3.22]{Horn} is invoked once again. Since Eq.~\eqref{eq:H} can be written as
\begin{equation*}
    \mathcal{M} = \begin{bmatrix}e_1^{}\boldsymbol{a}+e_3^{}\boldsymbol{c} &e_2^{}\boldsymbol{a}+e_1^*\boldsymbol{c} \end{bmatrix} \begin{bmatrix}\boldsymbol{a}^H \\ \boldsymbol{c}^H\end{bmatrix} ,
\end{equation*}
its two non-zero eigenvalues are the same as eigenvalues of the matrix written below
\begin{equation*}
    \begin{bmatrix}\boldsymbol{a}^H \\ \boldsymbol{c}^H\end{bmatrix} \begin{bmatrix}e_1^{}\boldsymbol{a}+e_3^{}\boldsymbol{c} &e_2^{}\boldsymbol{a}+e_1^*\boldsymbol{c} \end{bmatrix} = \begin{bmatrix}e_1^{} &e_2^{} \\ e_3^{} &e_1^*\end{bmatrix} .
\end{equation*}
After performing several algebraic manipulations, the eigenvalues of $\mathcal{M}$ are obtained as 
\begin{equation}
    \begin{aligned}
    \lambda_{\ell}(x_m^{}) &= \mathrm{Re}\left\{\frac{\gamma}{2} \left(  e^{j \phi} \eta_1^{}\eta_2^* |x_m^{}|^2 + \eta_2^* x_m^* \right)\right\} \\ &
    \begin{aligned}
      ~~+(-1)^{\ell} \Bigg[&\mathrm{Re}\Big\{\frac{\gamma}{2}\left(  e^{j \phi} \eta_1^{}\eta_2^* |x_m^{}|^2 + \eta_2^* x_m^* \right)\Big\}^2 \\
         &
         \begin{aligned}
         +\frac{\gamma}{4} \bigg(&1 + \Delta  |x_m^{}|^2 +  e^{ j \phi}\eta_1^{} x_m^{} \\
         & + e^{-j \phi}  \eta_1^*x_m^* + \frac{1}{\gamma}\bigg)\Bigg]^{\frac{1}{2}}.
         \end{aligned}
    \end{aligned}
    \end{aligned}
     \label{eq:realEigenvalues}
\end{equation} 
together with $N_r-2$ zeros. It is worth to mention that $|\eta_1| \ll 1$ and $|\eta_2| \ll 1$ such that the eigenvalues expressed in Eq.~\eqref{eq:realEigenvalues} are dominated by the third component, which yields that $\lambda_1(x_m)$ is negative and $\lambda_2(x_m)$ is positive. With these two eigenvalues, the testing statistic in Eq.~\eqref{eq:indefiniteGQF} is an indefinite quadratic form of Gaussian random variables whose distribution is given in~\cite{Al-Naffouri2016}. For the studied case, the obtained testing statistic follows asymmetrical Laplace distribution (ALD), of which CDF and PDF are as in Eq.~\eqref{eq:CDF_real} and Eq.~\eqref{eq:PDF_real}, respectively. 

Let us denote two components of the eigenvalues, for simplicity, as
\begin{equation*}
\begin{aligned}
\varepsilon_m^{} &= \mathrm{Re}\left\{\frac{\gamma}{2}\eta_2^{}\left(  e^{j \phi}\eta_1^{} |x_m^{}|^2 + x_m^* \right)\right\},\\ A_m^{} &=  \frac{\gamma}{4} \left(1 + \Delta  |x_m^{}|^2 +  e^{ j \phi}\eta_1^{} x_m^{} +e^{-j \phi}  \eta_1^*x_m^* + \frac{1}{\gamma} \right) .
\end{aligned}
\end{equation*}
Hence, the eigenvalues can be rewritten as 
\begin{equation}
    \lambda_{\ell}(x_m) = \varepsilon_m^{} +(-1)^{\ell} \sqrt{\varepsilon_m^2 +A_m^{}}, \quad\ell \in\{ 1,2\}.
\end{equation}
Since $\eta_1^{}$ and $\eta_2^{}$ are much less than 1, it can be seen that  $A_m^{} \gg \varepsilon_m^{}$. In addition, $\varepsilon_m^{}$ is dominated by $\mathrm{Re}\{\gamma\eta_2^{}x_m^*/2\}$, and $A_m^{}$ is dominated by the $\gamma/4$ term which is not rely on $x_m^{}$ so $A_0^{}$ and $A_1^{}$ have similar values. 

The expectation and variance of the ALD are
\begin{subequations}
\begin{gather}
    \mathrm{E}\{\zeta|x_m\} = -\lambda_1(x_m^{}) -\lambda_2(x_m^{}) = -2\varepsilon_m^{} ,\\
    \mathrm{Var}\{\zeta|x_m\} =\lambda_1^2(x_m^{}) + \lambda_2^2(x_m^{})  = 4 \varepsilon_m^2 + 2A_m^2 ,
\end{gather}
\end{subequations}
respectively.
Therefore, when the tag adopt BPSK modulation, we have $ \mathrm{E}\{\zeta|x_0\}\approx - \mathrm{E}\{\zeta|x_1\}$ and $ \mathrm{Var}\{\zeta|x_0\} \approx  \mathrm{Var}\{\zeta|x_1\}$.



\ifCLASSOPTIONcaptionsoff
  \newpage
\fi

\bibliographystyle{IEEEtran}

\bibliography{AmBC}

\end{document}